\pdfoutput=1
\documentclass{llncs}
\usepackage[utf8]{inputenc}
\title{Data-Driven Robust Control for Type 1 Diabetes Under Meal and Exercise Uncertainties}

\author{
Nicola Paoletti\inst{1} \and 
  Kin Sum Liu\inst{1} \and
  Scott A. Smolka\inst{1} \and
 Shan Lin\inst{2}} 
\institute{Department of Computer Science, Stony Brook University, USA
\and
Department of Electrical and Computer Engineering, Stony Brook University, USA
}

\usepackage{cite}
\usepackage{graphicx}
\usepackage[usenames,dvipsnames]{xcolor}
\usepackage[cmex10]{amsmath}
\usepackage{bm}
\usepackage{amssymb}
\usepackage{url}
\usepackage{textcomp}
\usepackage{algorithm}
\usepackage{rotating}
\usepackage{wrapfig}

\usepackage[noend]{algpseudocode}

\makeatletter
\def\BState{\State\hskip-\ALG@thistlm}
\makeatother

\usepackage{array}
\usepackage{caption}
\usepackage[font=footnotesize,caption=false]{subfig}
\usepackage{fixltx2e}
\usepackage{lipsum}
\usepackage{multirow}

\usepackage{url}

\usepackage{todonotes}

\renewcommand{\paragraph}[1]{\vspace*{4pt}\noindent\textit{#1:}}
\newcommand{\predh}{{N_p}}
\newcommand{\ctrlh}{{N_c}}

\newcommand{\review}[1]{#1}

\begin{document}

\maketitle


\begin{abstract}
We present a fully closed-loop design for an artificial pancreas (AP) which regulates the delivery of insulin for the control of Type~I diabetes.  Our AP controller operates in a fully automated fashion, without requiring any manual interaction (e.g.~in the form of meal announcements) with the patient. 
A major obstacle to achieving closed-loop insulin control is the uncertainty in those aspects of a patient's daily behavior that significantly affect blood glucose, especially in relation to meals and physical activity. To handle such uncertainties, we develop a data-driven robust model-predictive control framework, where we capture a wide range of individual meal and exercise patterns using uncertainty sets learned from historical data. These sets are then used in the controller and state estimator to achieve automated, precise, and personalized insulin therapy. We provide an extensive \textit{in silico} evaluation of our robust AP design, demonstrating the potential of this approach, without explicit meal announcements, to support high carbohydrate disturbances and to regulate glucose levels in large clusters of virtual patients learned from population-wide survey data. 
\end{abstract}

\section{Introduction}\label{sect:intro}
\emph{Type~1 diabetes} (T1D) is an autoimmune disease where the pancreas is not able to autonomously produce a sufficient amount of insulin to regulate blood glucose (BG) levels, thereby inhibiting glucose uptake in muscle and adipose (fatty) tissue. In healthy subjects, pancreatic $\beta$ cells are responsible for the release of insulin in amounts commensurate with current BG levels. This regulation maintains healthy BG values within tight ranges, normally between 70-200 mg/dL. 
In T1D, T cell–mediated destruction of insulin-producing $\beta$ cells occurs, leading to high BG levels. 

In the U.S.\ alone, more than 29 million people suffer from diabetes, among which about 5\% have T1D \cite{centers2014national}. T1D patients need to wear an insulin pump for the injection of \textit{basal} and \emph{bolus} insulin.  Basal insulin is a low and continuous dose that covers insulin needs outside meals. Bolus insulin is a single high dose for covering meals. 

The concept of closed-loop control of insulin, a.k.a.\ the artificial pancreas (AP), involves a continuous glucose monitor (CGM) that provides glucose measurements (with a typical period of $5$ minutes) to a control algorithm running inside the insulin pump or on a peripheral device (e.g.~smartphone or tablet) connected to the pump \cite{zavitsanou2016embedded}.  The controller adjusts the insulin therapy to maintain healthy BG levels and to avoid \textit{hyperglycemia} (BG above the healthy range) as well as \textit{hypoglycemia} (BG below the healthy range). AP systems have been extensively studied in the last 20 years \cite{hovorka2011closed}, but only lately cleared for clinical trials \cite{ly2015day,kovatchev2017feasibility} and commercialization. 

The recently FDA-approved MINIMED 670G by Medtronic\footnote{\scriptsize \url{https://www.medtronicdiabetes.com/products/minimed-670g-insulin-pump-system}} is the first commercial AP system, and can regulate the basal insulin rate automatically. It is referred to as a ``hybrid closed-loop" device as patients need to manually announce the amount of carbohydrate (CHO) and time of each meal to receive the appropriate bolus insulin dose. This manual procedure is a burden to the patient and inherently dangerous as incorrect information can lead to incorrect insulin dosage and, in turn, harmful BG levels.

While meals are the major source of uncertainty in BG control, another important factor is physical activity, which accelerates glucose absorption and thus requires a reduced insulin dosage. To build fully automated \textit{closed-loop} AP systems, it is essential to design insulin control algorithms that are \textit{robust} to the patient’s behavior and activities. 



In this paper, we propose a \textit{data-driven, robust model-predictive control} (robust MPC) framework for the \text{closed-loop} control of insulin administration, both basal and bolus, for T1D patients under uncertain meal and exercise events. Such a framework seeks to eliminate the need for meal announcements by the patient, to fully automate insulin regulation. We capture the wide range of individual meal and exercise patterns using \textit{uncertainty sets} learned from historical data.

Following~\cite{bertsimas2013data}, we construct uncertainty sets from data so that they cover the underlying (unknown) distribution with prescribed probabilistic guarantees. Leveraging such information, our robust MPC system computes the insulin administration profile that minimizes the worst-case performance with respect to these uncertainty sets, so providing a principled way to deal with uncertainty. 

Besides uncertainty, another challenging aspect of closed-loop control is \emph{state estimation}, which is needed to recover the full state of the model (used within MPC) from CGM measurements. Not only are these measurements noisy and delayed with respect to BG (the CGM detects glucose in the interstitial fluid), but we also need to estimate, along with the state, current meal and exercise uncertainties. 

For this purpose, we designed a moving-horizon state estimator (MHE) \cite{rao2003constrained,gondhalekar2014moving,lee2014design} that, similar to MPC, exploits a prediction model to find the most likely state estimate given the observations. Crucially, data-driven uncertainty sets improve the estimation by constraining the admissible meal and exercise uncertainties.


To the best of our knowledge, our robust MPC design for an AP is the first approach to leverage data-driven techniques to enhance robust insulin control and state estimation, supporting at the same time both meal and exercise uncertainties.  In summary, our main contributions are the following.
\begin{itemize}
\item We formulate a closed-loop AP design based on robust MPC to optimize BG levels under meal and exercise uncertainties.
\item We apply data-driven techniques to construct uncertainty sets that provide probabilistic guarantees on the robust MPC solution.
\item We design an MHE that leverages data to make informed estimates for BG and uncertainty parameters.
\item We provide an extensive \textit{in-silico} evaluation of our design, including one-meal simulations, one-day high carbohydrate intake scenarios, and one-day simulations of large clusters of virtual patients learned from population-wide survey data sets (CDC NHANES). 
\item Overall, our robust closed-loop AP is able to keep BG within safe levels between 84\% and 100\% of the time, outperforming an implementation of a hybrid closed-loop AP and state-of-the art robust control algorithms~\cite{szalay2014linear}.
\end{itemize}
\section{System Overview}\label{sec:system}

\begin{wrapfigure}{r}{0.6\textwidth}
\centering
\includegraphics[width=.6\columnwidth]{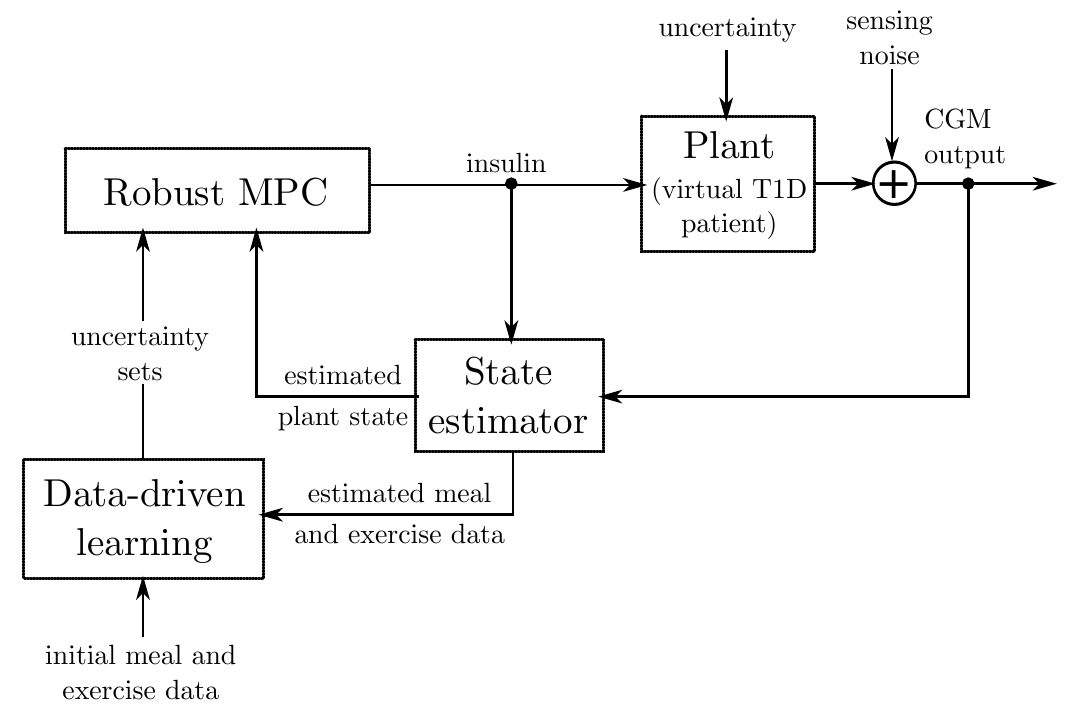}
\caption{Robust artificial pancreas design.}
\label{fig:system_design}
\vspace{-0.3em}
\end{wrapfigure}

The design of our proposed data-driven robust artificial pancreas is illustrated in Figure \ref{fig:system_design}. The \textit{robust MPC} component (described in Section \ref{sec:rob_mpc}) is responsible for computing the insulin administration strategy (both basal and bolus) that optimizes, over a finite time horizon, the predicted BG profile against worst-case realizations of the uncertainty parameters, used to capture unknown meal and exercise information. 

Uncertainty sets describe the domains of the uncertainty parameters and are derived by the \textit{data-driven learning} component (see Section \ref{sec:usets}), starting from a dataset 
about the patient's meal and exercise schedules. Uncertainty sets can be also updated online as new data (estimated or announced) comes along, in this way enabling the continuous learning of the patient's behavior.

\setlength{\intextsep}{5pt}%
\setlength{\columnsep}{8pt}

At this stage, we analyze our robust artificial pancreas design \textit{in silico}. Thus, the \textit{plant} is given by a system of differential equations (see Section \ref{sec:model}) describing the gluco-regulatory dynamics of a virtual T1D patient, as well as the effects of insulin and random disturbances (i.e.~unknown realizations of the uncertainty parameters). 

In order to faithfully reproduce real-life settings, we assume that the state of the plant (BG) cannot be observed by the controller, but that we can only access (noisy) CGM measurements. We designed a \textit{moving-horizon state estimator} (described in Section \ref{sec:state_est}) that, based on a bounded history of CGM measurements and estimations, computes the most likely plant state. Importantly, this component also provides estimates for the uncertainty parameters, which can be used to update the uncertainty sets.

\section{Plant Model}
\label{sec:model}
\subsection{Uncertainty parameters}
\label{sec:uncertain}
To account for uncertainty in meal consumption, we consider the parameter $D_G^t$, which describes the \textit{rate of CHO ingestion} at time $t$. As in the exercise model of \cite{lenart2002modeling,hernandez2008extension,jacobs2015incorporating,resalat2016design}, physical activity is represented by parameters $\mathit{MM}^t$, the \textit{percentage of active muscular mass} at time $t$, and $\mathit{O2}^t$, the \textit{percentage of maximum oxygen consumption} which can be combined to reproduce arbitrary kinds of physical activity.

$\mathit{MM}^t$ corresponds to the ratio between the active muscular mass and the total muscular mass, with typical values being $\mathit{MM}^t = 0\%$ at rest and $\mathit{MM}^t = 25\%$ for a two-legged exercise. $\mathit{O2}^t$ describes the oxygen consumed relative to the maximum oxygen consumption of the subject, and thus, represents a subject-independent measure of exercise workload. As in \cite{lenart2002modeling,hernandez2008extension}, typical values are $8\%$ at rest, $30\%$ for light activity, $60\%$ for moderate activity, and $90\%$ for intense activity. In our scenario, these meal and exercise parameters are not observed or measured, and are thus represented by an uncertainty parameter vector $\mathbf{u}^t = (D_G^t,\mathit{MM}^t,\mathit{O2}^t)$. The effects of these parameters on blood glucose are described in Section~\ref{sec:patient}, in which the patient's gluco-regulatory model is presented.

\subsection{Patient Model}
\label{sec:patient}
We consider the nonlinear ODE gluco-regulatory model of Jacobs et al.~\cite{jacobs2015incorporating,resalat2016design}, which extends Hovorka's well-established model \cite{hovorka2004nonlinear,wilinska2005insulin,wilinska2010simulation} to capture the effect of exercise on BG. The model describes the dynamics of glucose and insulin in the human body, i.e., their absorption, metabolism, excretion and transport between compartments (tissues and organs). In addition to insulin, Jacobs' model also allows for the automated control of glucagon, i.e.~the hormone antagonistic to insulin that protects against hypoglycemia. In our work, however, we leave aside glucagon. 
Model parameters (listed in Table~\ref{tbl:model_params} of the appendix) 
are deterministic and represent the physiological characteristics (e.g.~transport or consumption rates) of a single virtual subject. 

At time $t$, the inputs to the system are the subcutaneous insulin infusion rate, $\iota^t$ (mU/min), and the uncertainty parameter values, $\mathbf{u}^t = (D_G^t,\mathit{MM}^t,\mathit{O2}^t)$. The output corresponds to the CGM measurement. The state-space representation of the system is as follows:
\begin{align}
\dot{\mathbf{x}}(t) = & {\bf F}\left(\mathbf{x}(t), \iota^t, \mathbf{u}^t \right)
\label{eq:state_space_form}\\
y(t) = & h\left(\mathbf{x}(t)\right) + v^t \label{eq:output}
\end{align}
where $\mathbf{x}$ 
is the 14-dimensional state vector that evolves according to the ODE system $\bf{F}$, which is given below 
(see Appendix \ref{app:full_ODE} for the full set of equations). 
Eq.~\ref{eq:output} describes the CGM measurement $y$, which is derived from $\mathbf{x}$ with the measurement model $h$ and subject to an additive measurement noise $v^t \in \mathcal{N}(0,q^t)$, where $q^t$ is the noise variance.  We fix $q^t = 0.1521$ mmol$^{2}$/L$^{2}$ constant for all $t$, corresponding to a standard deviation equal to 5\% of the ideal glucose value.

\begin{wrapfigure}{r}{0.6\textwidth}
\centering
\includegraphics[width=.6\textwidth, trim=4cm 15cm .5cm 3cm, clip=true]{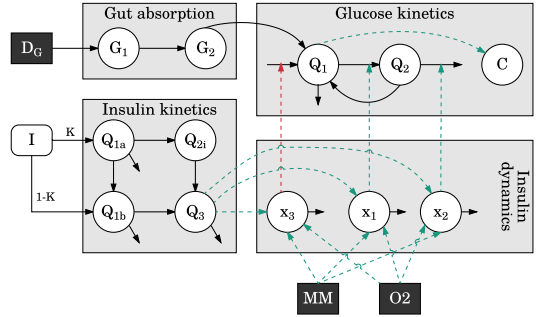}
\caption{Schema of the gluco-regulatory ODE system and its four main subsystems. White circles: ODE variables; black boxes: uncertainty parameters; white rounded box: insulin input; solid black arrows: flows of glucose or insulin; dashed green/red arrows: positive/negative interactions between variables.}\label{fig:ODE_schema}
\vspace{-0.3em}
\end{wrapfigure}

Figure \ref{fig:ODE_schema} illustrates a high-level schema of the ODE system ${\bf F}$. The \textit{gut absorption} subsystem \cite{wilinska2010simulation} uses a chain of two compartments, $G_1$ and $G_2$ (mmol), to describe digestion of ingested CHO, given by the uncertainty parameter $D_G^t$. 

The \textit{glucose kinetics} subsystem describes the glucose masses in the accessible (where BG measurements are made) and non-accessible compartments, respectively through variables $Q_1$ and $Q_2$ (mmol).  BG concentration, $G$ (mmol/L), is the main variable we aim to control, and is derived from $Q_1$ as $G(t) = Q_1(t)/V_G$, where $V_G$ is the glucose distribution volume. Variable $C$ is the glucose concentration in the interstitial fluid, which has a delayed response w.r.t.\ the concentration in the blood $G$. $C$ corresponds to the glucose detected by the CGM sensor and thus, the measurement function $h$ of Eq.~\ref{eq:output} maps the state vector ${\bf x}(t)$ to $C(t)$.

\setlength{\intextsep}{5pt}%
\setlength{\columnsep}{8pt}

The \textit{insulin kinetics} subsystem models the absorption of the fast-acting insulin $\iota^t$, i.e.~our control input (in mU/min), and its transport through compartments $Q_{1a}$, $Q_{1b}$, $Q_{2i}$ and $Q_{3}$ (in mU) \cite{wilinska2005insulin}. This model assumes a slow insulin absorption pathway consisting of compartments $Q_{1a}$ (subcutaneous insulin mass) and $Q_{2i}$ (non-accessible insulin), and a fast pathway that includes only $Q_{1b}$ (subcutaneous). $K$ represents the proportion in which the input insulin $\iota^t$ is distributed into the two pathways. $Q_3$ is the plasma insulin mass, from which we derive the plasma insulin concentration $I$ (mU/L) as $I(t) = Q_3(t)/V_I$, where $V_I$ is the insulin distribution volume.

The \textit{insulin dynamics} subsystem defines the effects of insulin on blood glucose through variables $x_1, x_2, x_3$. Variable $x_1$ (min$^{-1}$) promotes glucose distribution; $x_2$ (min$^{-1}$) promotes glucose disposal	; and $x_3$ (unitless) inhibits endogenous glucose production. The overall subsystem decrease blood glucose masses $Q_1$ and $Q_2$ and in turn, BG concentration $G$. Plasma insulin levels $I$ directly increase $x_1, x_2, x_3$. Uncertainty parameters $\mathit{MM}^t$ (active muscular mass) and $\mathit{O2}^t$ (target workload in terms of oxygen consumption) increase $x_1, x_2, x_3$ indirectly, through state variables $\mathit{UA}$ (mg/min) and $\mathit{O2}_m$ (unitless), not shown in the figure. They characterize physical activity and describe, respectively, the glucose uptake due to active muscular tissue and the actual percentage of maximum oxygen consumption. 

\paragraph{Initial conditions} The initial state of the system is derived at a steady-state BG level of 7.8 mmol/L \cite{szalay2014linear}, assuming no meal and exercise. We use a nonlinear equation solver (MATLAB's \texttt{fsolve}) to find $\mathbf{x}(0)$ and the basal insulin level $\bar{\iota}$ such that $\dot{\mathbf{x}}(0) = F\left(\mathbf{x}(0), \bar{\iota}, \mathbf{u}^0 \right) = \mathbf{0}$ (see Eq.~\ref{eq:state_space_form}), where the uncertainty parameters $\mathbf{u}^0$ are given by $D_G^0 = 0$, $\mathit{MM}^0 = 0$ and $\mathit{O2}^0 = 8$ (oxygen consumption at rest). 
Following~\cite{jacobs2015incorporating}, we further assess the physiologic feasibility of the initial conditions by checking that: 1)~in absence of insulin, steady-state BG is above 300 mg/dL, and 2)~delivery of high-dose insulin (15 U/h) results in a steady-state BG below 100 mg/dL.

\section{Robust MPC}\label{sec:rob_mpc}
Since we want to optimize the BG profile against worst-case realizations of the uncertainty parameters, at each time step $t$, the robust MPC computes the insulin infusion $\iota^t$ as the solution of the following non-linear minimax optimization problem:

{\vspace*{-20pt}\footnotesize
\begin{align}
& \underset{\iota^t,\ldots,\iota^{t+N_c-1}}{\min}\underset{\mathbf{u}^t,\ldots,\mathbf{u}^{t+N_p-1}}{\max} \sum_{k=1}^\predh  d(\tilde{\mathbf{x}}(t+k)) + \beta\cdot \sum_{k=0}^{\ctrlh - 1} (\Delta \iota^{t+k})^2 & \label{eq:rob_mpc_prob}
\end{align}
\vspace*{-15pt}
\begin{align}
\text{subject to: } & \iota^{t+k} \in D_{\iota} & (k=0,\ldots,N_c -1) \label{eq:rob_mpc2}\\
& \iota^{t+k} = \bar{\iota}  & (k=N_c,\ldots,N_p -1)\label{eq:rob_mpc4}\\
& \mathbf{u}^{t+k} \in \mathcal{U}^{t+k} &  (k=0,\ldots,N_p -1)\label{eq:rob_mpc3}\\
& \tilde{\bf x}(t) = \hat{\bf x}(t) &\label{eq:rob_mpc5}\\
& \dot{\tilde{\mathbf{x}}}(t+k) = F(\tilde{\bf x}(t+k),\iota^{t+k},\mathbf{u}^{t+k}) & (k=0,\ldots,N_p -1)\label{eq:rob_mpc6}
\end{align}}\noindent
where $N_c$ and $N_p$ are the control and prediction horizon (in minutes), respectively; constraint (\ref{eq:rob_mpc2}) states that the control input $\iota$ must belong to some set $D_{\iota}$ of admissible insulin infusion rates; through (\ref{eq:rob_mpc4}), we impose that $\iota$ is fixed to the basal insulin rate $\bar{\iota}$ outside the control horizon; (\ref{eq:rob_mpc3}) states that, at any time point $t+k$ in the prediction horizon, uncertainty parameters $\mathbf{u}^{t+k}$ must belong to the corresponding uncertainty sets $\mathcal{U}^{t+k}$; constraint (\ref{eq:rob_mpc5}) and (\ref{eq:rob_mpc6}) restrict how the robust MPC computes the predicted state vector $\tilde{\bf x}$: for the initial state, it uses the estimated plant state at time $t$, $\hat{\bf x}(t)$, while following states are predicted using the same plant model (see Equation \ref{eq:state_space_form}). We set control and prediction horizons to $N_c = 100$ min and $N_p = 150$ min, respectively, as opposed to \cite{resalat2016design} where $N_c = 20$ and $N_p = 200$: preliminary experiments suggested that large $N_p$ values and small $N_c$ values cause excessive insulin therapy and hypoglycemia.

We design the cost function so as to optimize the following two objectives: 
\begin{enumerate}
\item Minimize the sum of squared distances between the predicted BG level $\tilde{\bf x}_G(t+k)$ and a target trajectory $R(t+k)$:

{\vspace*{-10pt}\footnotesize
\begin{equation}\label{eq:cost1}
d(\tilde{\mathbf{x}}(t+k)) = \gamma(t+k) \cdot \left( \tilde{\bf x}_G(t+k) - R(t+k)\right)^2 
\end{equation}
}\noindent
where $\gamma(t+k) = \gamma$ if $\tilde{\bf x}_G(t+k) < R(t+k)$ and $1$ otherwise. (Remind that ${\bf x}_G(t) = G(t) = Q_1(t)/V_G$ in the glucose kinetics subsystem) Parameter $\gamma \geq 1$ allows defining asymmetric cost functions where predicted BG values below the target are penalized more than those above the target. 
Glucose control is naturally asymmetric given that hypoglycemia leads to more severe consequences than (temporary) hyperglycemia, and, as shown in \cite{gondhalekar2016periodic}, asymmetric costs effectively contribute avoiding hypoglycemia. 
\item Minimize step-wise changes in the control input $(\Delta \iota^{t+k})^2$, where $\Delta \iota^{t+k} = \iota^{t+k} - \iota^{t+k-1}$, and $\iota^{t-1}$ corresponds to the control input in the previous iteration, or to the basal insulin rate $\bar{\iota}$ if $t=0$. 
\end{enumerate}
In our setup, we fix the target trajectory to $R(t+k) = 7.8$ mmol/L for all time instants and set penalty $\beta$ to $1/50$. We set the asymmetric cost penalty to $\gamma = 2$, after experimenting with different values 
(see Appendix \ref{sect:asymm_costs}). 

\paragraph{Optimization algorithm} We solve problem (\ref{eq:rob_mpc_prob}) using non-linear optimization techniques, where, for a fixed control strategy $\iota^t,\ldots,\iota^{t+N_c-1}$, the objective function value is given in turn as the result of maximizing the objective function over the uncertainty parameters (and with fixed $\iota^t,\ldots,\iota^{t+N_c-1}$). To solve both minimization and maximization problems, we use MATLAB's \texttt{fmincon}. 
To reduce the computational cost of this optimization method, we decrease the number of decision variables by assuming that, in the prediction model, control inputs change with period $10$ min, and uncertainty parameters with period $30$ min. 

%
\paragraph{Hybrid closed-loop (HCL) variant} To compare with our robust MPC approach, we develop a hybrid closed-loop insulin pumps where only basal insulin is automatically regulated and the patient is responsible for bolus insulin. This reduces to a MPC that has no knowledge of meals and exercise, and thus, approximates the behavior of a current state-the-art approved device that requires explicit meal announcement. In our settings, this is equivalent to fixing the uncertainty parameters to their default values at rest.

Then the optimization problem of the  HCL controller reduces to:

{\vspace*{-15pt}\footnotesize
\begin{align}
& \underset{\iota^t,\ldots,\iota^{t+N_c-1}}{\min} \sum_{k=1}^\predh  d(\tilde{\mathbf{x}}(t+k)) + \beta\cdot \sum_{k=0}^{\ctrlh - 1} (\Delta \iota^{t+k})^2 &\\
&\text{subject to } (\ref{eq:rob_mpc2}, \ref{eq:rob_mpc4}, \ref{eq:rob_mpc5}, \ref{eq:rob_mpc6}) \text{ and } \mathbf{u}^{t+k} = (0,0,8) \quad (k=0,\ldots,N_p -1).&\nonumber
\end{align}}\noindent
Note that the constraints on the insulin therapy are the same of the robust controller (\ref{eq:rob_mpc2}-\ref{eq:rob_mpc4}) meaning that the HCL controller is free to synthesize bolus-like therapy profiles too. This will also serve as the baseline controller in the evaluation part of Section \ref{sec:results}.

\subsection{State estimation}\label{sec:state_est}
This component allows to recover an estimate of the current state, which is used in the following iteration by the robust MPC as the initial state for its predictions (see Eq.~\ref{eq:rob_mpc5}). 
Following \cite{rao2003constrained,haseltine2005critical}, we designed a moving-horizon state estimator (MHE) that works in a finite-horizon fashion similar to an MPC problem, and allows estimating the current state starting from previous estimations and a bounded history of observed CGM measurements.

For an estimation window of size $N$, MHE is based on simulating a model of the plant from time $t-N$ to $t$ and aims at finding the  model trajectory ${\bf x}(t-N), \ldots {\bf x}(t)$ that minimizes the discrepancies between simulated and estimated states, and between simulated and measured outputs (CGM).  Then, $\hat{\bf x}(t)$ is chosen as the final state of the optimal trajectory. 


Crucially, our estimator also works as a meal and physical activity detector \cite{dassau2008detection,lee2009closed,weimer2016physiology}: in addition to the plant state, we compute the most likely sequence of uncertainty parameters $\mathbf{u}^{t-N},\ldots, \mathbf{u}^t$, corresponding to decision variables in our optimization problem as they are inputs of the model. The MHE problem boils down to the following non-linear optimization problem:

{\vspace*{-10pt}\footnotesize
\begin{align}
& \underset{{\bf x}(t-N), \ldots {\bf x}(t), \mathbf{u}^{t-N},\ldots, \mathbf{u}^t}{\min} \ \mu \cdot \|{\bf x}(t-N) - \hat{\bf x}(t-N)\|^2 + \sum_{k=0}^{N-1} \frac{\|v^{t-k}\|^2}{q^{t-k}}   & \label{eq: MHE}
\end{align}
\vspace*{-15pt}
\begin{align}
\text{subject to: } & v^{t-k} = y(t-k) - h({\bf x}(t-k)) & (k=N-1,\ldots,0) \label{eq:mhe2}\\
& \dot{{\mathbf{x}}}(t-k) = F({\bf x}(t-k),\iota^{t-k},\mathbf{u}^{t-k}) & (k=N,\ldots,0)\label{eq:mhe3}\\
& \mathbf{u}^{t-k} \in \mathcal{U}^{t-k} &  (k=N,\ldots,0)\label{eq:mhe4}
\end{align}}\noindent
where (\ref{eq:mhe2}) defines the measurement discrepancy $v^{t-k}$ at time ${t-k}$ as the difference between the measured and simulated output, $y({t-k})$ and $h({\bf x}(t-k))$, respectively (see also Eq.~\ref{eq:output}); and (\ref{eq:mhe3}) states that $\mathbf{x}$ evolves according to the same ODE model of the plant, with $\iota^{t-k}$ being the insulin input previously computed by the robust MPC. We remark that data-driven uncertainty sets play an important role also in state estimation, since they constrain the domain of the corresponding estimated uncertainty parameters, as per (\ref{eq:mhe4}). The problem is solved using MATLAB's \texttt{fmincon} non-linear solver.

The first addend of the cost function penalizes the discrepancy between the initial state of the simulated trajectory and the corresponding state estimation, where $\mu > 0$ is a weighting factor. The second addend penalizes measurement discrepancies, weighted by the inverse of the measurement noise variance $q^{t-k}$ (see Eq.~\ref{eq:output}). 
In the original formulation of the MHE \cite{rao2003constrained,haseltine2005critical}, the cost function includes discrepancies for all the states in the trajectory. Our simplification comes from the fact that we do not consider random noise in the model (but only in the measurements), and thus, the trajectory ${\bf x}(t-N), \ldots, {\bf x}(t)$ is fully determined by the initial state ${\bf x}(t-N)$ and by the uncertainty parameters $\mathbf{u}^{t-N},\ldots, \mathbf{u}^t$. Further, this greatly improves computational efficiency because variables ${\bf x}(t-N+1), \ldots, {\bf x}(t)$ are strictly constrained by the ODE in Eq.~(\ref{eq:mhe3}).  \review{In practice, this means that the decision variables reduce to ${\bf x}(t-N), \mathbf{u}^{t-N},\ldots, \mathbf{u}^t$. }


The MHE has an important probabilistic interpretation: when $N=t$ (unbounded horizon), the MHE problem corresponds to maximizing the joint probability for the trajectory of states ${\bf x}(t-N), \ldots, {\bf x}(t)$ given the measurements $y(t-N),\ldots, y(t)$ \cite{rao2003constrained}. 

\subsection{Building data-driven uncertainty sets}\label{sec:usets} 


In this section, we describe how to build the uncertainty sets used within the robust MPC and the state estimator to restrict the domain of the admissible meal and exercise parameters. We apply the approach of \cite{bertsimas2013data} where the authors present a general schema for designing uncertainty sets from data for robust optimization (of which robust MPC is an instance). The key idea is to define an uncertainty set that captures possible realizations of the uncertain parameters and then optimize against worst-case realizations within this set. Importantly, this method requires no information about the underlying distribution of the parameters and provides a probabilistic guarantee (an upper bound) on the likelihood that the true realized cost is higher than the optimal `worst-case' cost computed by the robust controller.


Let us characterize an uncertainty set $\mathcal{U}$ by means of a so-called robust constraint $f(\mathbf{u},\mathbf{x}) \leq 0$, where $\mathbf{u}$ is the uncertainty parameter and $\mathbf{x}$ is the optimization variable, corresponding in our case to the state vector plus insulin input. Recall that the true distribution $\mathbb{P}^*$ of $\mathbf{u}$ is unknown. Given confidence level $\epsilon > 0$,  $\mathcal{U}$ should satisfy two conditions: (1) the robust constraint $f$ is computationally tractable. (2) \textit{$\mathcal{U}$ implies a probabilistic guarantee for $\mathbb{P}^*$ at level $\epsilon$}, that is, for any solution $\mathbf{x}^* \in \mathbb{R}^k$ and for any function $f(\mathbf{u}, \mathbf{x})$ concave in $\mathbf{u}$ for all $\mathbf{x}$,

{\vspace*{-7pt}\footnotesize
\begin{equation*}\label{eq:uset_guarantee}
\text{if }  f(\mathbf{u}, \mathbf{x}^*) \leq 0\ \forall \mathbf{u} \in \mathcal{U}, \text{then } \mathbb{P}^*(f(\mathbf{u},\mathbf{x^*}) \leq 0) \geq 1-\epsilon.
\end{equation*}
}\noindent
The data-driven schema we follow is based on sampling a set of data points $\mathcal{S}$ i.i.d. from the true distribution $\mathbb{P}^*$ and uses hypothesis testing to construct the uncertainty sets with such guarantees. In particular, for confidence level $\alpha<1$, the schema employs the corresponding $(1-\alpha)$ confidence region to build $\mathcal{U}$. With the proper construction, the following theorem from \cite[Sect.\ 3.2]{bertsimas2013data} holds.

\begin{theorem}
With probability at least $1-\alpha$ with respect to the sampling, the resulting set $\mathcal{U}(\mathcal{S}, \epsilon, \alpha)$ implies a probabilistic guarantee at least $\epsilon$ for $\mathbb{P}^*$.
\end{theorem}

In \cite{bertsimas2013data}, the authors show how different uncertainty sets are built depending on the assumptions about $\mathbb{P}^*$, and, in turn, on the suitable statistical test. In this work we consider box sets (i.e.~multi-dimensional intervals), which make no assumptions on $\mathbb{P}^*$ and are suitable for data with missing values. 
The assumptions and the full construction are described in Appendix B. 
The application of other types of uncertainty sets, able for instance to capture temporal dependencies and correlation between meals and exercise, is in our future plans.

To shrink the size of uncertainty set, we employ the following two strategies: 
1) prior to set construction, we classify the input data and partition it into a number of clusters so as to obtain tighter sets and more customized, patient-specific control strategies. 
2) based on Algorithm 1 of \cite{bertsimas2013data}, we use bootstrapping \cite{efron1994introduction} to approximate the threshold of the test statistics, by estimating the sampling distribution of the statistics through re-sampling with replacement. 

We remark that the construction of uncertainty sets is performed off-line and thus has no computational footprint on the robust controller.

\section{Results and Discussion}
\label{sec:results}
We evaluate our robust control algorithm through a number of experiments for simulating:  intake of a single meal (Section \ref{sect:one_meal_experiments}), exercise (Section \ref{sect:res_exercise}), one-day meal intake scenario with patient behavior learned from population-wide survey data (Section \ref{sect:NHANES}), and two-day scenario with irregular meal timing and unusually high CHO intake (Section \ref{sect:high_carbs}). Section \ref{sect:res_estimation} is dedicated to the analysis of state estimation. For each experiment, we compare the robust controller with the non-robust, hybrid closed-loop (HCL) variant introduced in Section \ref{sec:rob_mpc}.  
We also report the ideal performance by running a so-called \textit{perfect controller}, that can access both the full plant state (i.e.~does not need state estimation) and the exact values of the uncertainty parameters in the plant.

\paragraph{Hardware and performance}
We ran the experiments on a Windows~8 machine with an Intel Core~i7 processor and 32GB of DDR3 memory. We used MATLAB version 2016b. With this configuration, the average time to compute the insulin therapy over all the experiments ranged from 4 to 18 seconds, which is well within the CGM measurement period of 5 minutes.  This means that the controller works faster than real-time.  Given the significant performance improvement of modern embedded and mobile devices, we expect our algorithm to perform similarly as well once deployed on  such hardware platforms.

\paragraph{Performance indicators} To measure the efficacy of our robust controller design over multiple runs, we consider the following indicators:
\begin{itemize}
\item $t_{\sf < 3.9}$, $t_{\sf 3.9-11.1}$, $t_{\sf > 11.1}$: mean percentage of time spent in, respectively, hypoglycemia (BG $< 3.9$ mmol/L), normal ranges (BG between $3.9$ and $11.1$), and hyperglycemia (BG $> 11.1$). Clearly, we wish to maximize $t_{\sf 3.9-11.1}$ and minimize the other two indicators, keeping in mind that we can tolerate some temporary postpandrial hyperglycemia while hypoglycemia should be avoided as much as possible.
\item $BG_{\min}$, $BG_{\max}$: average low BG level and peak BG level, respectively, in mmol/L. An effective robust controller should keep  $BG_{\min}$ and $BG_{\max}$ as close as possible to the target BG level.
\item $\sum \iota$: mean total non-basal insulin (in U). This indicator measures the amount of insulin injected by the controller in order to cover meals, and thus excludes the contribution of basal insulin.
\end{itemize}

\noindent To evaluate state estimation, we further consider indicators $E_{D_G}$, $E_{MM}$, $E_{O2}$, i.e.~the mean absolute error between plant and estimated uncertain variable values, and $E_{BG}$, the mean absolute error between plant BG and estimated BG. 

\subsection{One-meal experiments}\label{sect:one_meal_experiments}

\begin{figure}
\centering
\newcommand{\relHeight}{.18}
\begin{footnotesize}
\begin{tabular}{cccc}
& {1) Meals As Expected} & {2) Outliers} & {3) Late Meals}\\
& \multicolumn{3}{c}{\includegraphics[width=.3\textwidth]{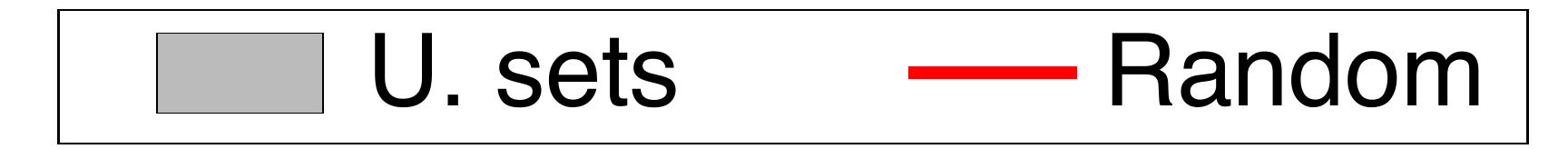}} \\[-.4cm]
\rotatebox{90}{$D_G$ (mmol/min)} &
\subfloat[]{\includegraphics[height=\relHeight\textwidth]{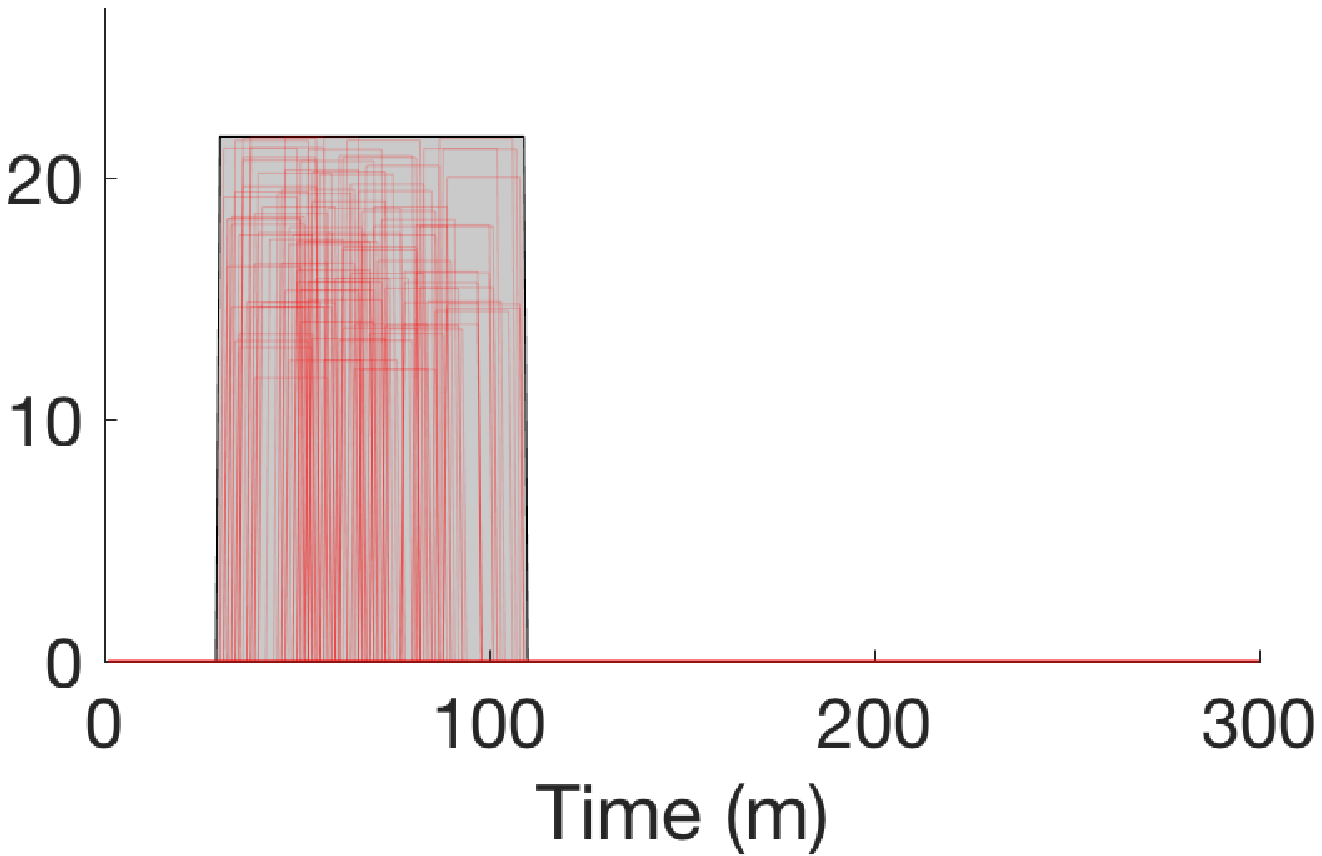}} &
\subfloat[]{\includegraphics[height=\relHeight\textwidth]{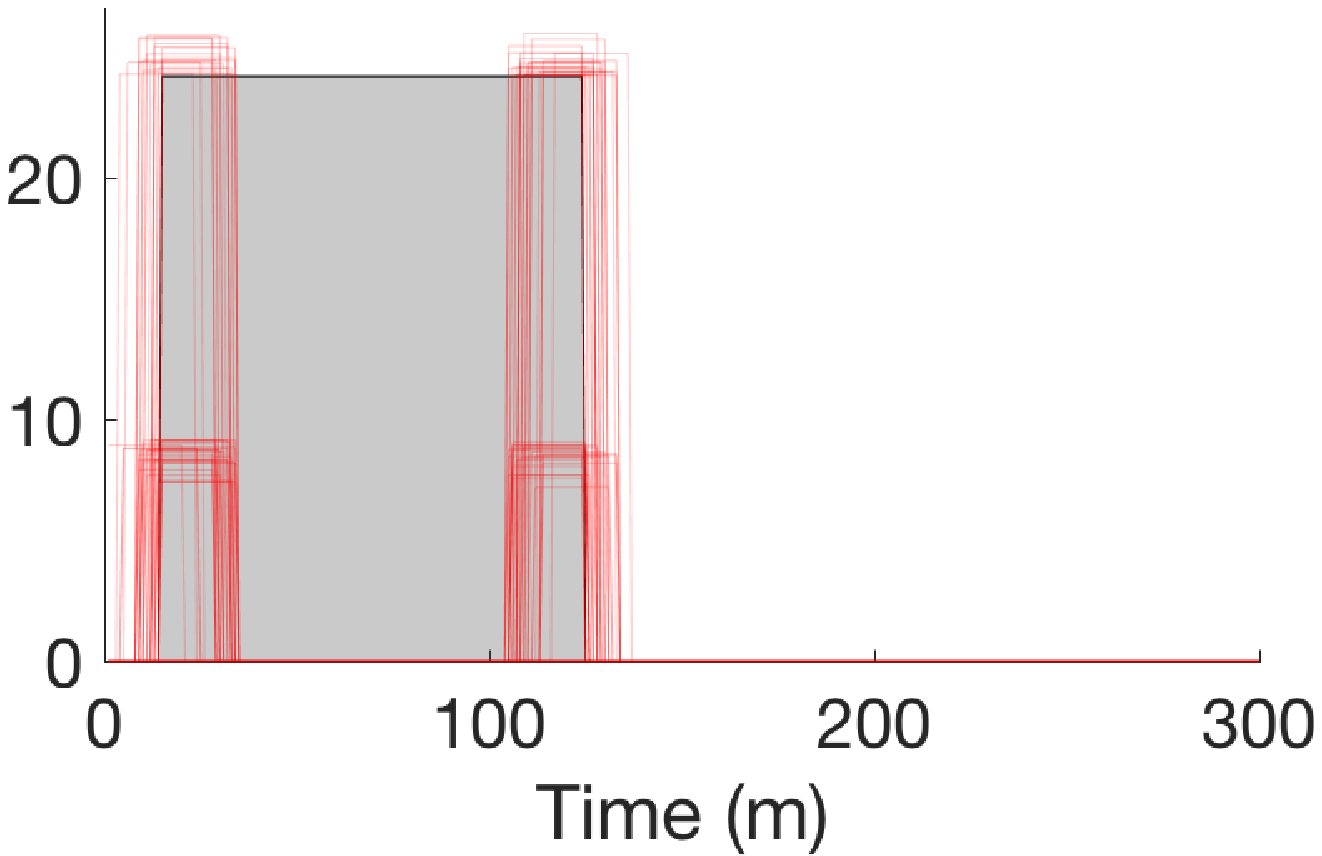}} &
\subfloat[]{\includegraphics[height=\relHeight\textwidth]{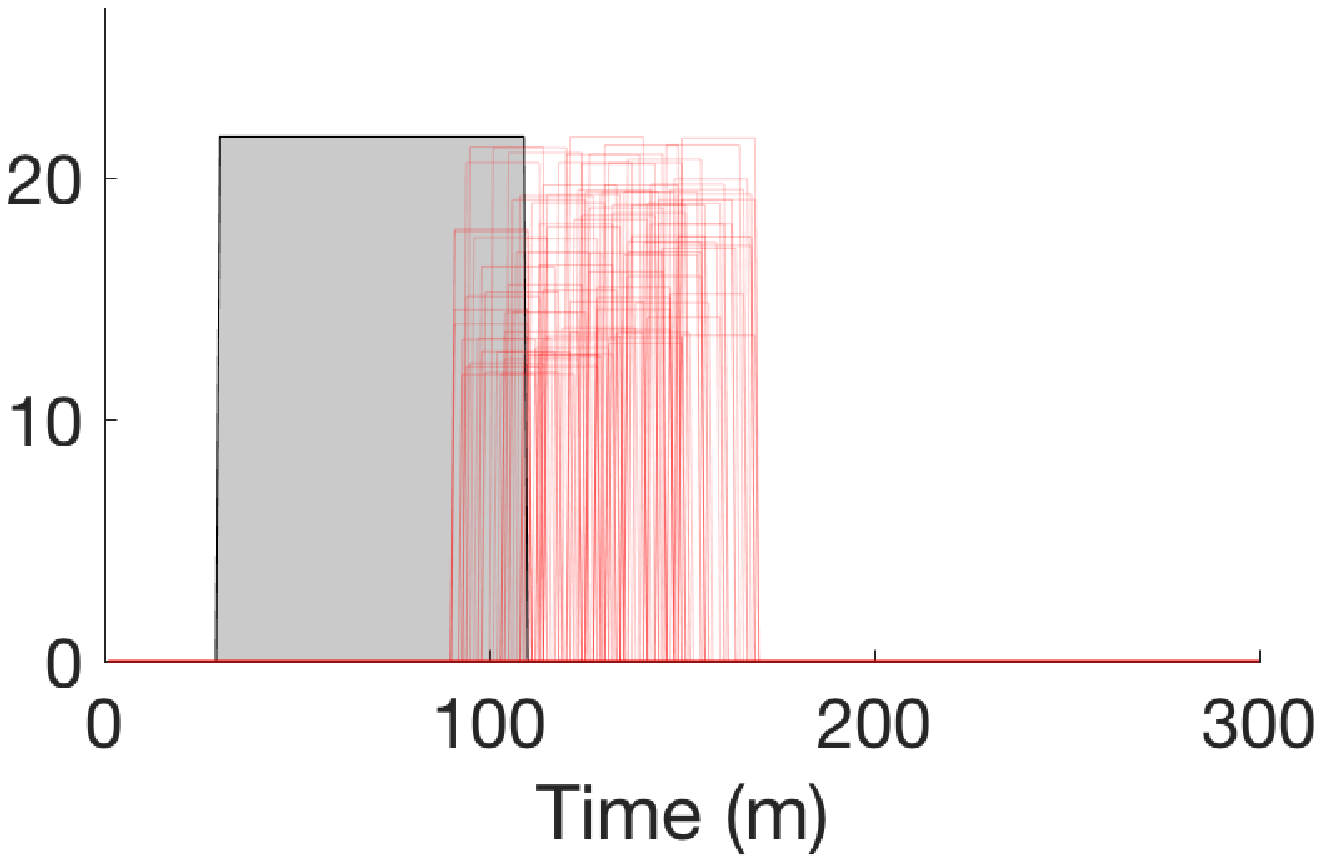}}\\
& \multicolumn{3}{c}{\includegraphics[width=.6\textwidth]{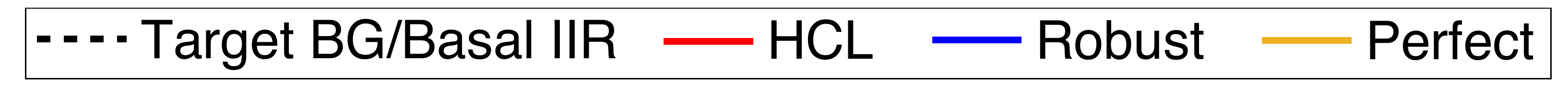}} \\[-.4cm]
\rotatebox{90}{{BG (mmol/L)}} &
\subfloat[]{\includegraphics[height=\relHeight\textwidth]{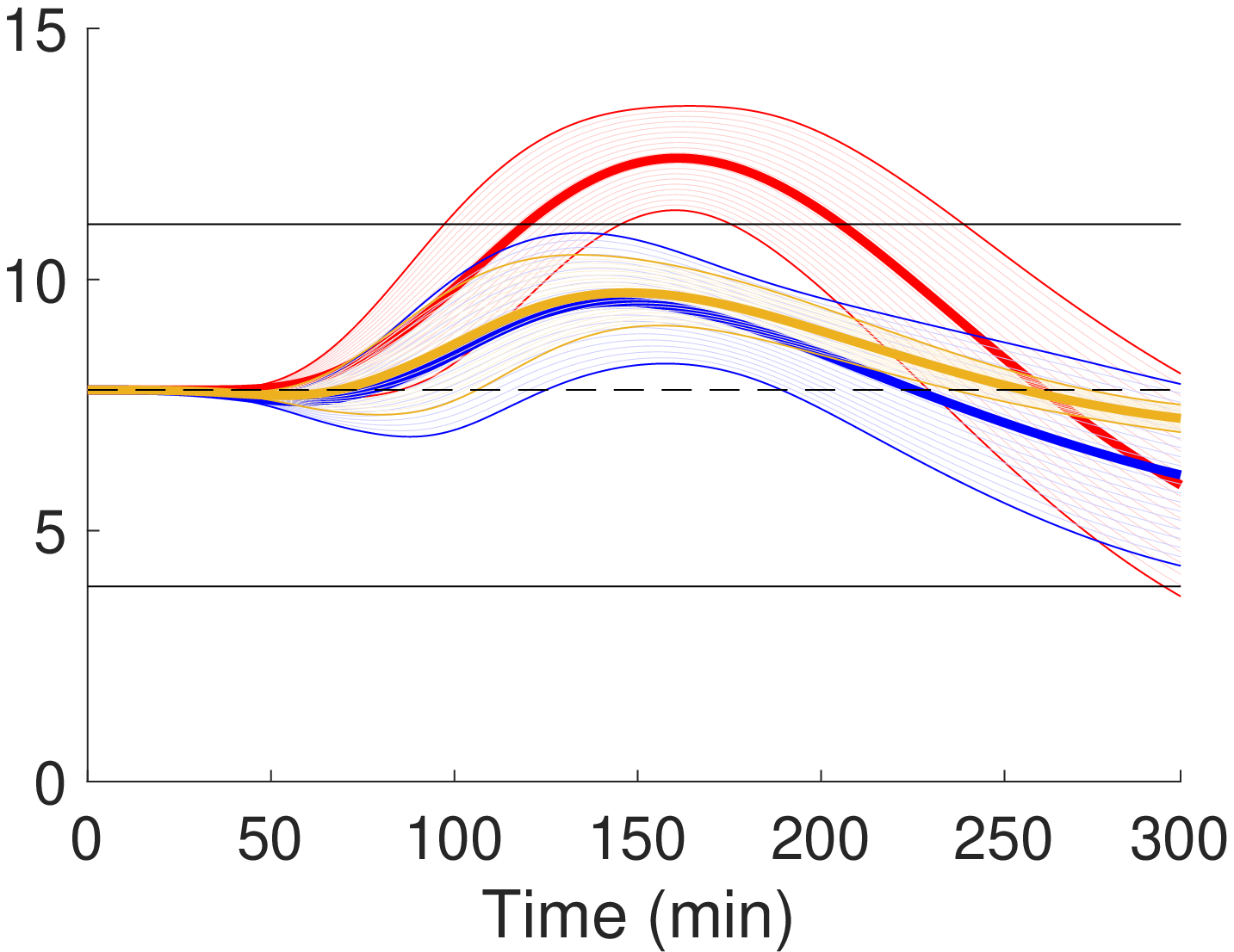}} &
\subfloat[]{\includegraphics[height=\relHeight\textwidth]{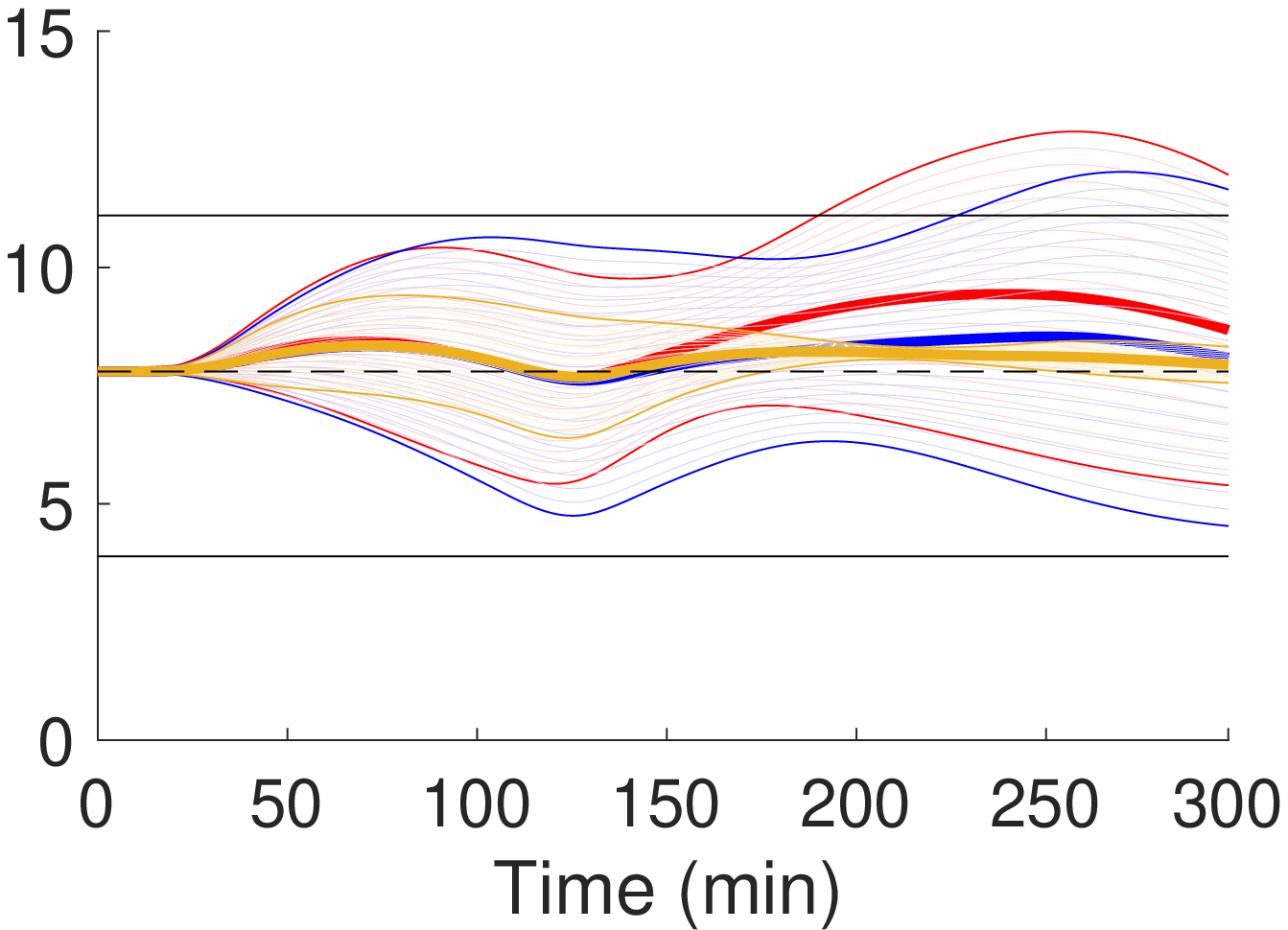}} & 
\subfloat[]{\includegraphics[height=\relHeight\textwidth]{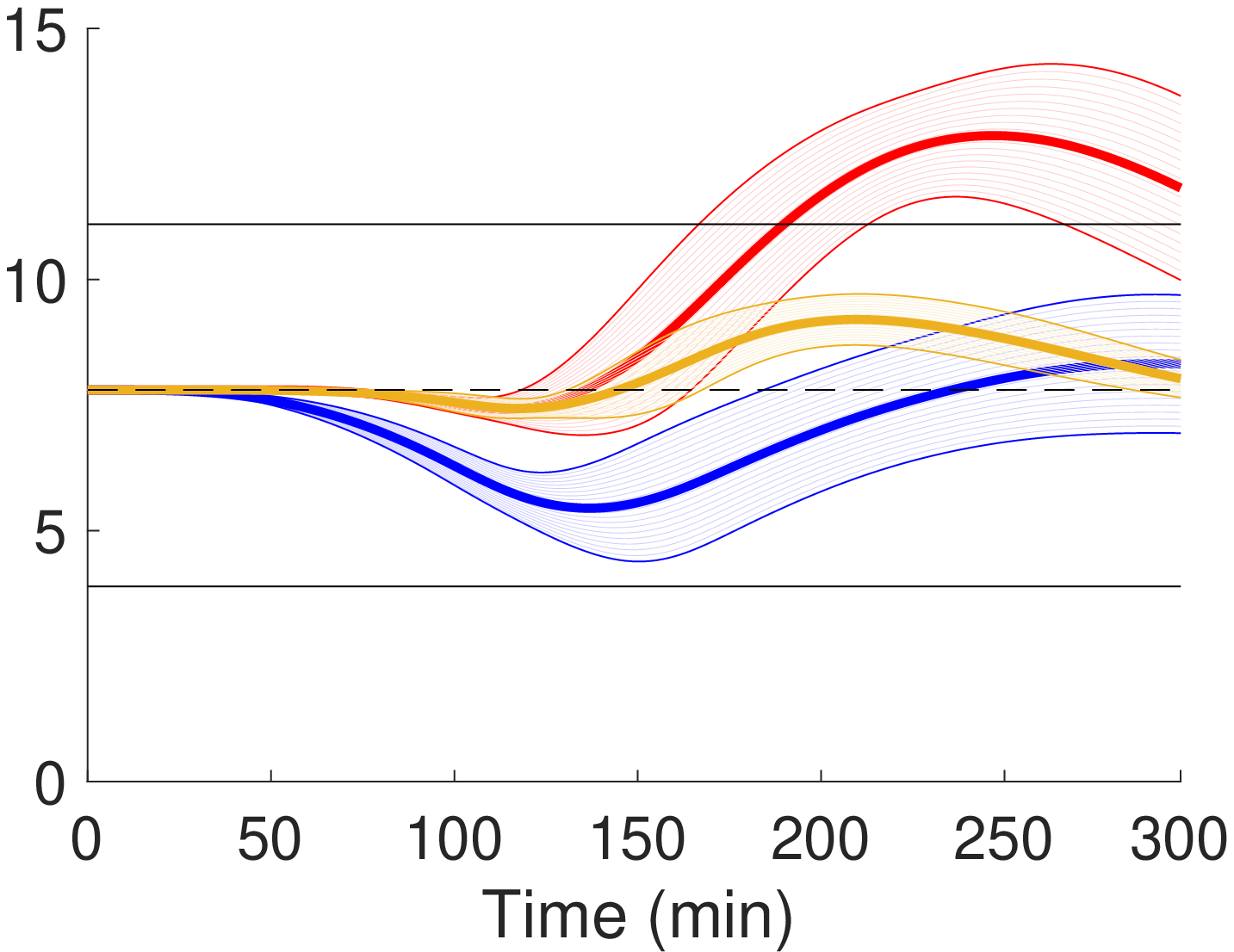}}\\[-.3cm]
\rotatebox{90}{{$\iota$ (mU/min)}} &
\subfloat[]{\includegraphics[height=\relHeight\textwidth]{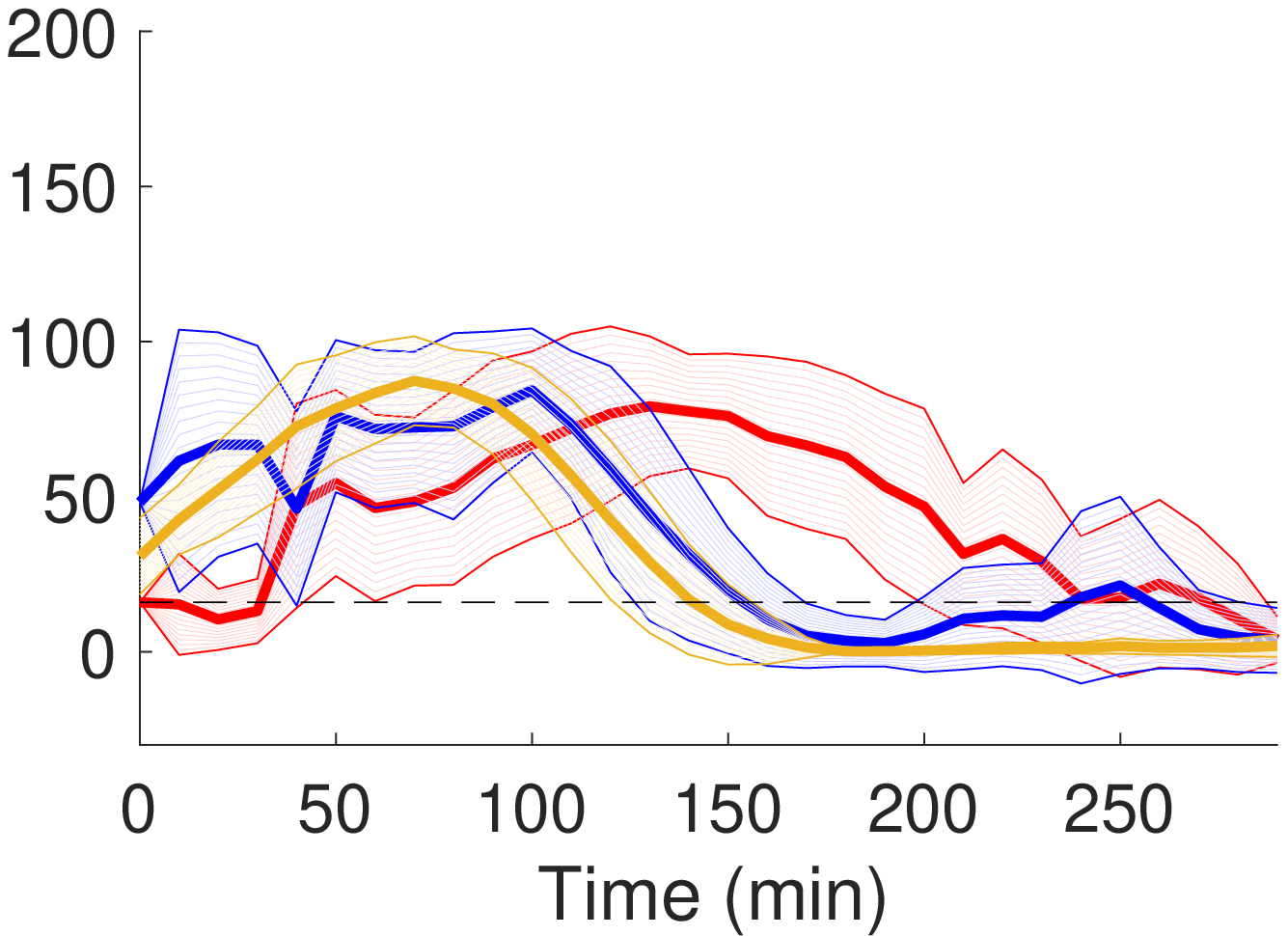}} &
\subfloat[]{\includegraphics[height=\relHeight\textwidth]{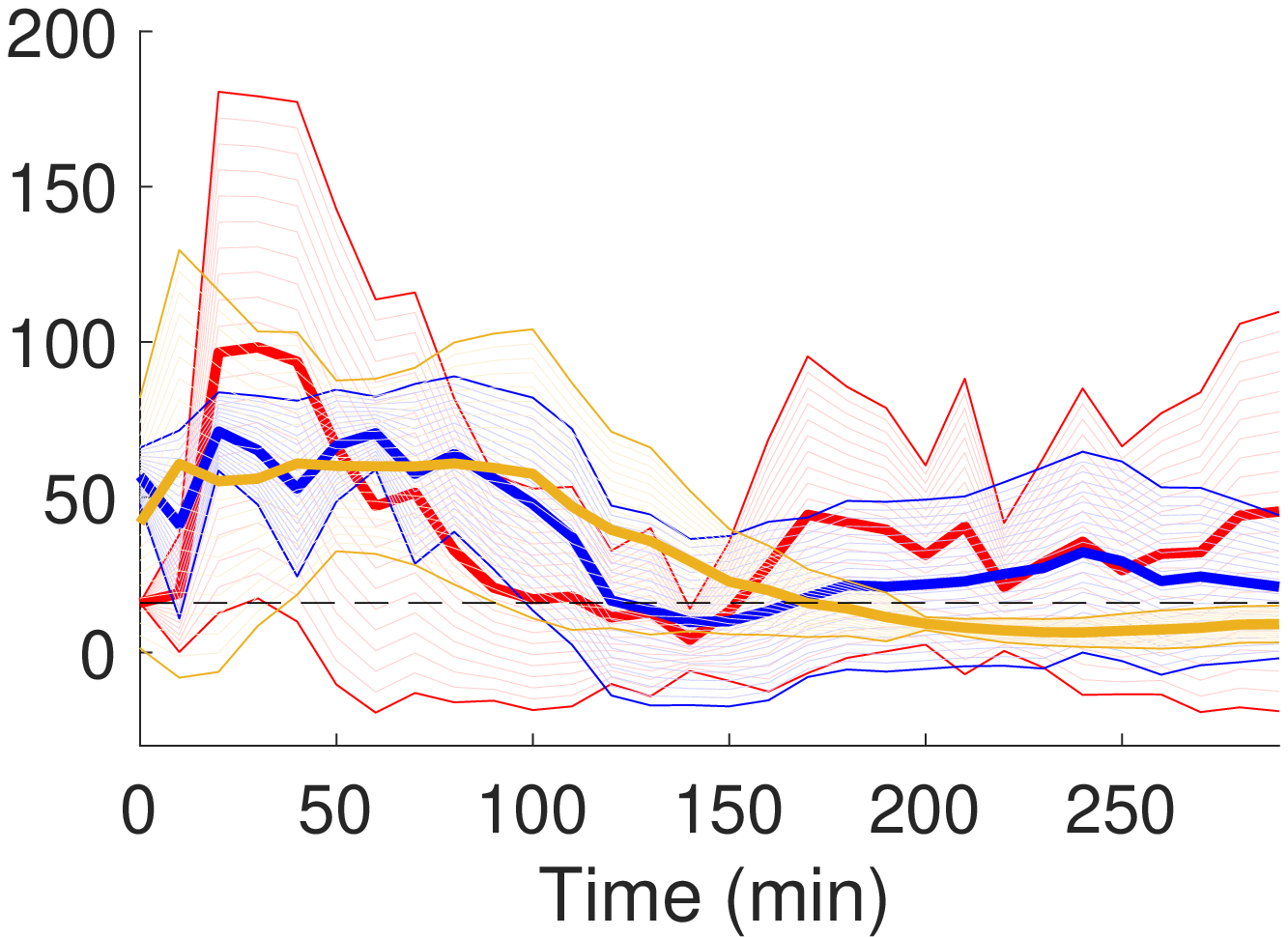}} & 
\subfloat[]{\includegraphics[height=\relHeight\textwidth]{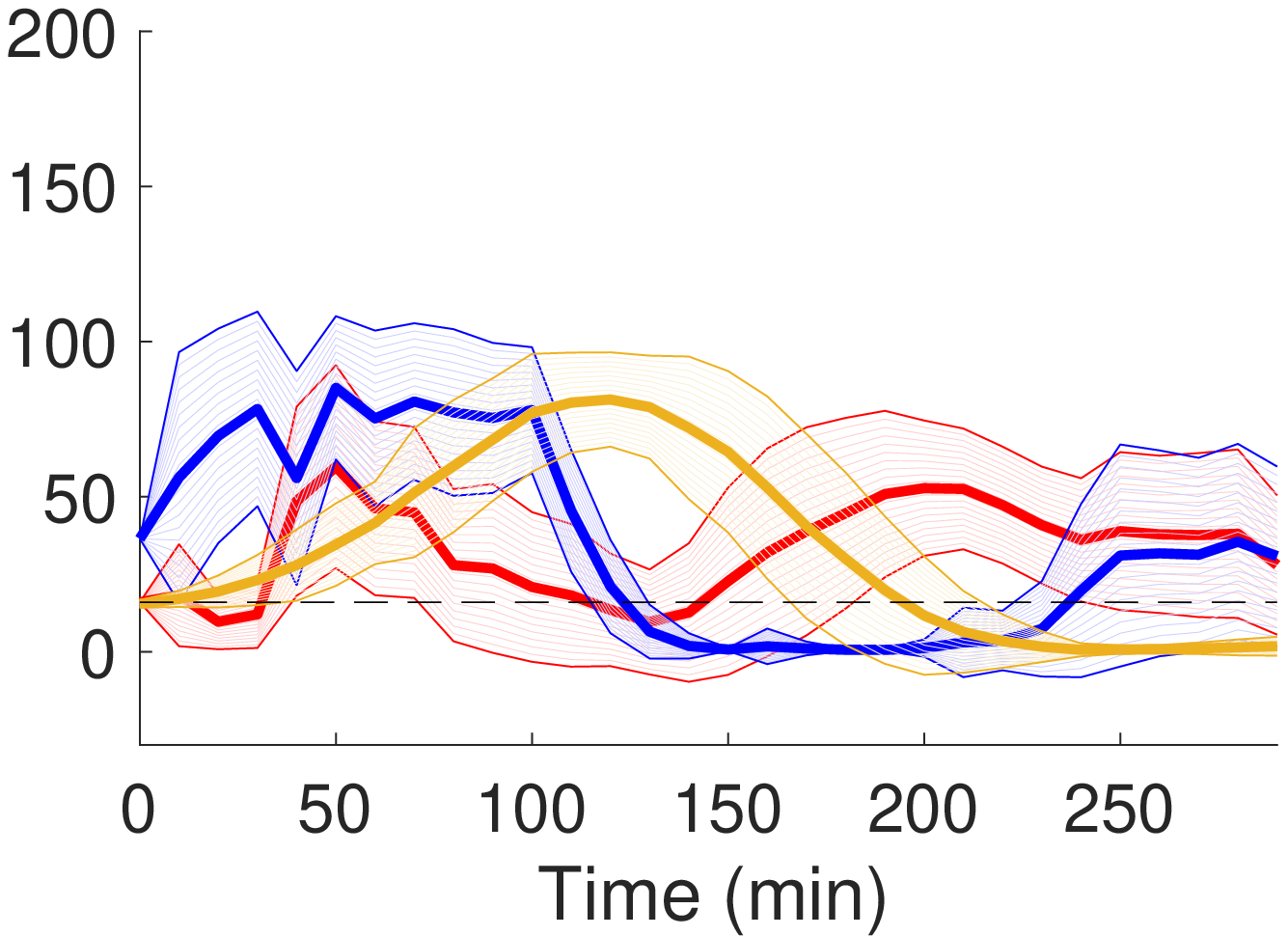}}
\end{tabular}
\end{footnotesize}
\\[10pt]
{\footnotesize
\begin{tabular}{r|cccccc}
& $t_{\sf < 3.9}$ & $t_{\sf 3.9-11.1}$ & $t_{\sf > 11.1}$ & $BG_{\min}$ & $BG_{\max}$  & $\sum \iota$\\ \hline
Scenario 1, perfect &  0\% &  99.69\% & 0.31\% & 7.15 & 9.91 & 4.38\\
Scenario 1, HCL &  1.6\% & 69.4\% & 29\% & \textbf{5.61}& 12.85 &  8.19\\ 
Scenario 1, robust &  \textbf{0.51\%} &\textbf{ 97.7\%} & \textbf{1.79\%} & 5.57 & \textbf{9.96} & 6.23\\ \hline
Scenario 2, perfect &  0\% &  100\% & 0\% & 7.03 & 8.84 &  4.67\\
Scenario 2, HCL &  1.03\% & 81.51\% & 17.45\% & \textbf{5.75}& 11.32 & 6.31\\ 
Scenario 2, robust & \textbf{0.28}\% & \textbf{84.19}\% & \textbf{15.53}\% & 5.16 & \textbf{10.94} & 5.82\\ \hline
Scenario 3, perfect &  0\% &  100\% & 0\% & 7.22 & 9.3 & 5.06\\
Scenario 3, HCL &  \textbf{0\%} & 67.25\% & 32.75\% & \textbf{7.19} & 13.34 & 5.05\\ 
Scenario 3, robust &  0.79\% & \textbf{99.03\%} & \textbf{0.18\%} & 5.09 & \textbf{8.77} &  5.64\\ \hline
\end{tabular}
}
\vspace*{-1em}
\caption{One-meal, 300-minute experiments (50 repetitions). \textbf{Top:} uncertainty sets and random realizations of parameter $D_G$ (rate of CHO ingestion). \textbf{Middle:} BG profiles (with solid black lines indicating the normal BG range). \textbf{Bottom:} synthetized insulin therapies. Thick solid lines indicate average BG/insulin values, and are surrounded by an area spanning $\pm$ 1 S.D. In the table, we highlight in bold the best value of each index between the robust and the HCL controllers. \vspace*{-2em}
}\label{fig:results_table_synth}
\end{figure}

We consider 300-minute simulations comprising a single meal, and three different synthetic scenarios (illustrated in Figure \ref{fig:results_table_synth} (a-c)), i.e.~where meals are sampled from arbitrary distributions. For each scenario and controller, we collect results for 50 repetitions. 
Details on the construction of uncertain sets from arbitrary distributions are given in Appendix \ref{sec:usets_short}. 

\paragraph{Scenario 1, meals as expected} in the uncertain plant, we assume a uniformly distributed meal with start time $t_s = \mathrm{unif}(30,90)$,  total amount of CHO (grams) $\mathsf{CHO} = \mathrm{unif}(42,78)$ and meal duration fixed to 20 minutes, during which CHO ingestion happens at a constant rate. Given that uniform distributions have bounded support, we can build tight box-type uncertainty sets (i.e.~intervals) that contain all possible realizations. 
This scenario allows us evaluating the adequacy of the controller when the plant behaves according to a known distribution, in other words, when we have accurate information for building uncertainty sets. 

\paragraph{Scenario 2, outliers} in this case, random meals behave as statistical outliers, i.e.~they are constantly distant from the expected value of the underlying distribution. To this purpose, we build the uncertainty sets 
under the assumption that meals are normally distributed with parameters $t_s = \mathcal{N}(60,15)$ and $\mathsf{CHO} = \mathcal{N}(60,9)$. The uncertainty sets are built so as to cover all possible realizations with z-score between -3 and 3 (i.e.~between -3 S.D. and +3 S.D. around the mean). However, to reproduce outliers, meals in the uncertain plant are sampled from the tails of the distributions (z-scores in $[-4,-3]$ and $[3,4]$). 

\paragraph{Scenario 3, late meals} here we consider the same settings as in Scenario 1, but with each random meal delayed of one hour. This models the situation where the controller has wrong information about the meal schedule, since it expects the meal to start, on the average, one hour earlier.

\vspace*{4pt}\noindent Results in Figure \ref{fig:results_table_synth} show that our robust controller attains very good performance, closely following the ideal behavior of the perfect controller in the first and third scenarios, where the virtual patient stays in normal ranges for $>$97\% of the time. In the outliers scenario, we register some postprandial hyperglycemia, because this scenario  is characterized by frequent high CHO intake. Overall, the robust controller is able to limit the time spent in hypoglycemia below $1$\% and consistently outperforms the HCL controller, staying in normal BG ranges for $3$\% to $31$\% more. 
Full statistics are reported in Table \ref{tbl:full_stat_one_meal} of the Appendix.

\subsection{Regulation during exercise}\label{sect:res_exercise}
We evaluate the behavior of the robust controller when the virtual patient is involved in physical activity, which, contrarily to meals, contributes to decreasing BG levels. We simulate a two-legged exercise consisting of two phases:
\begin{enumerate}
\item \textbf{Moderate activity}, with start time $t_s = \mathrm{unif}(40,80)$, duration $d = \mathrm{unif}(24,$ $36)$, active muscular mass  $\mathit{MM} = \mathrm{unif}(0.15,0.35)$, and oxygen consumption $\mathit{O2} = \mathrm{unif}(45,75)$; followed by
\item \textbf{Light activity}, where parameters stay as in the previous phase except for $\mathit{02} = \mathrm{unif}(15,45)$.
\end{enumerate}

\begin{figure}
\centering
\subfloat[$\mathit{MM}$]{\includegraphics[height=.2\textwidth]{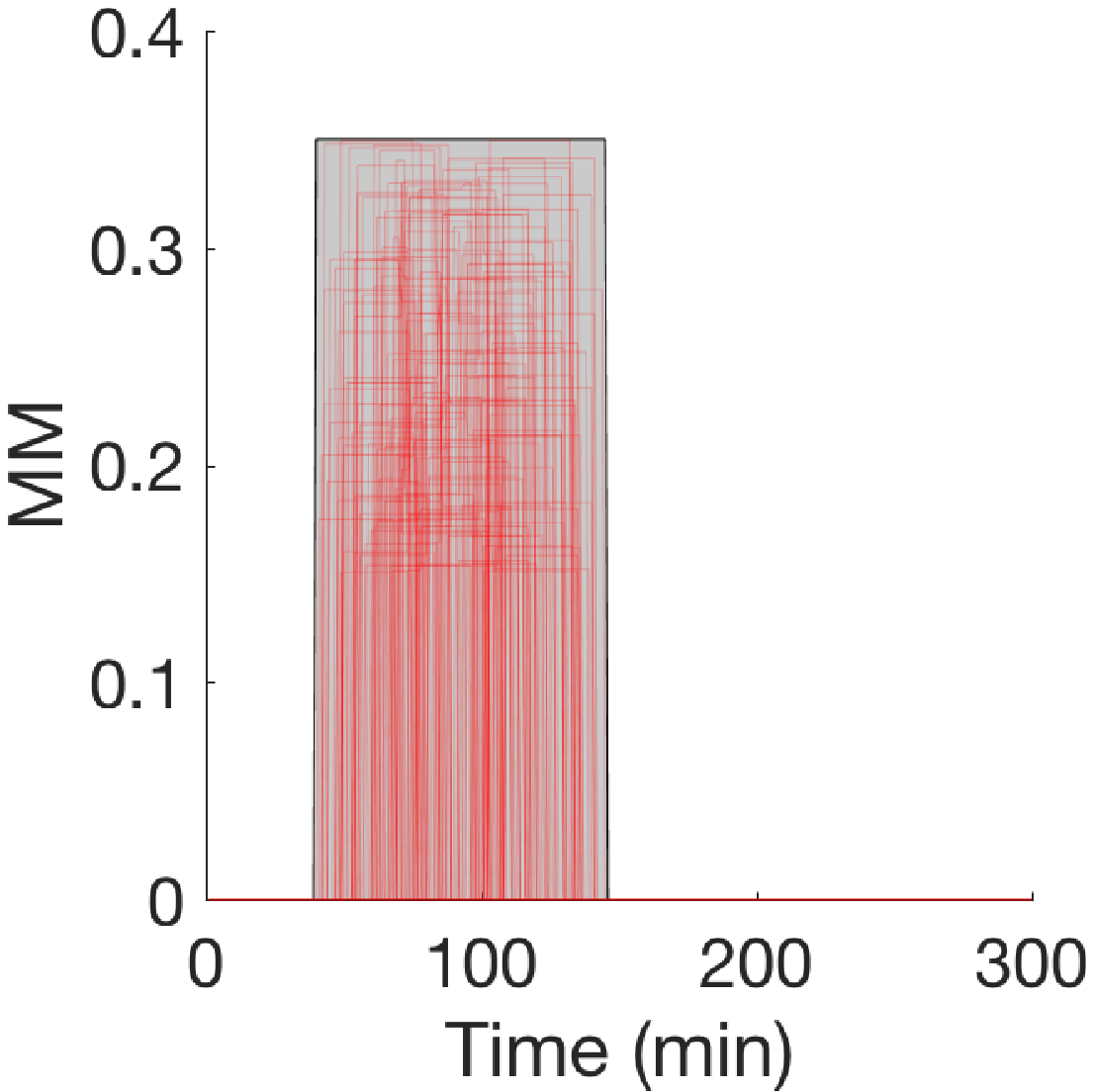}}
\hfill
\subfloat[$\mathit{O2}$]{\includegraphics[height=.2\textwidth]{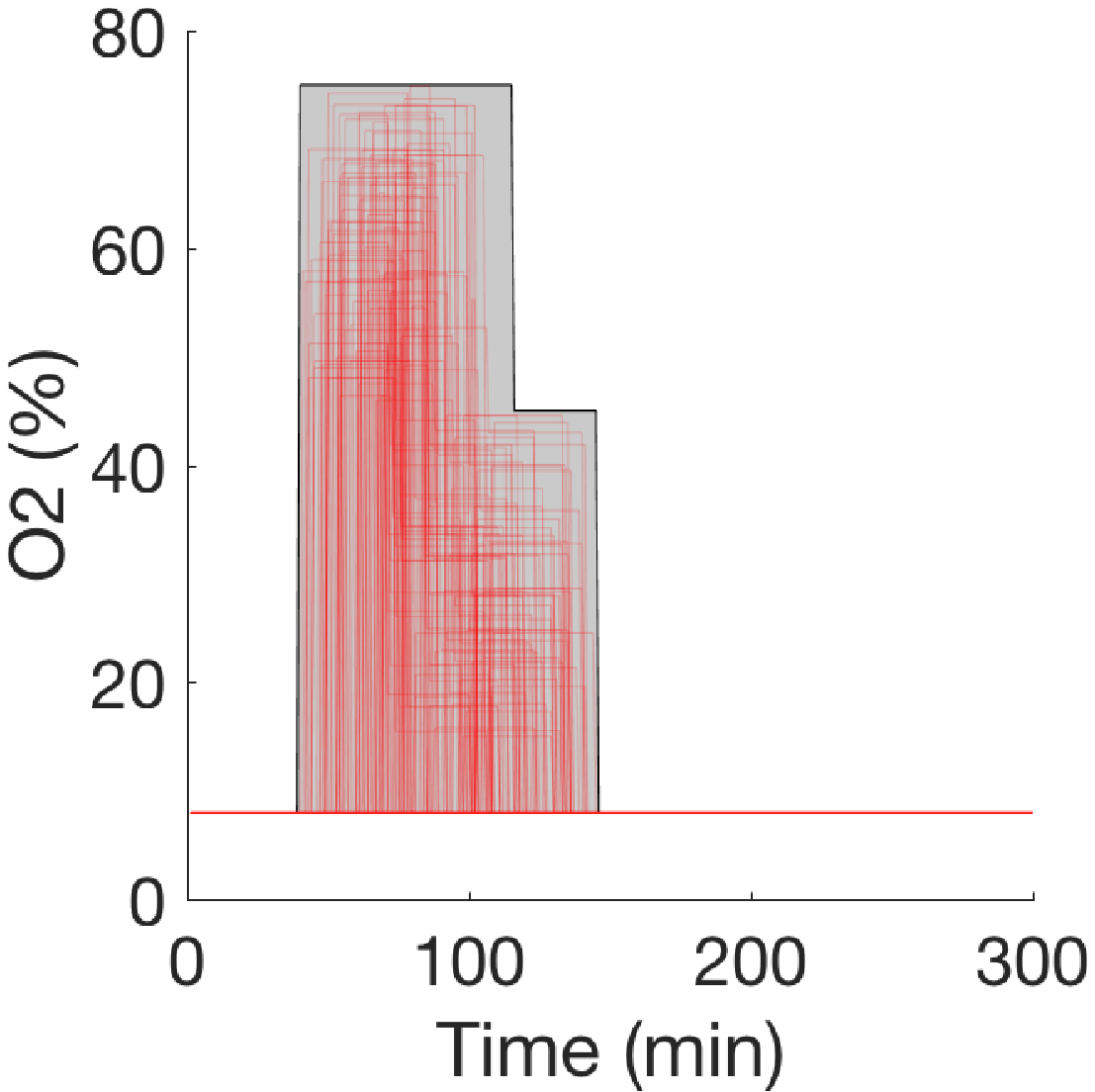}}
\hfill
\subfloat[BG]{\includegraphics[height=.2\textwidth]{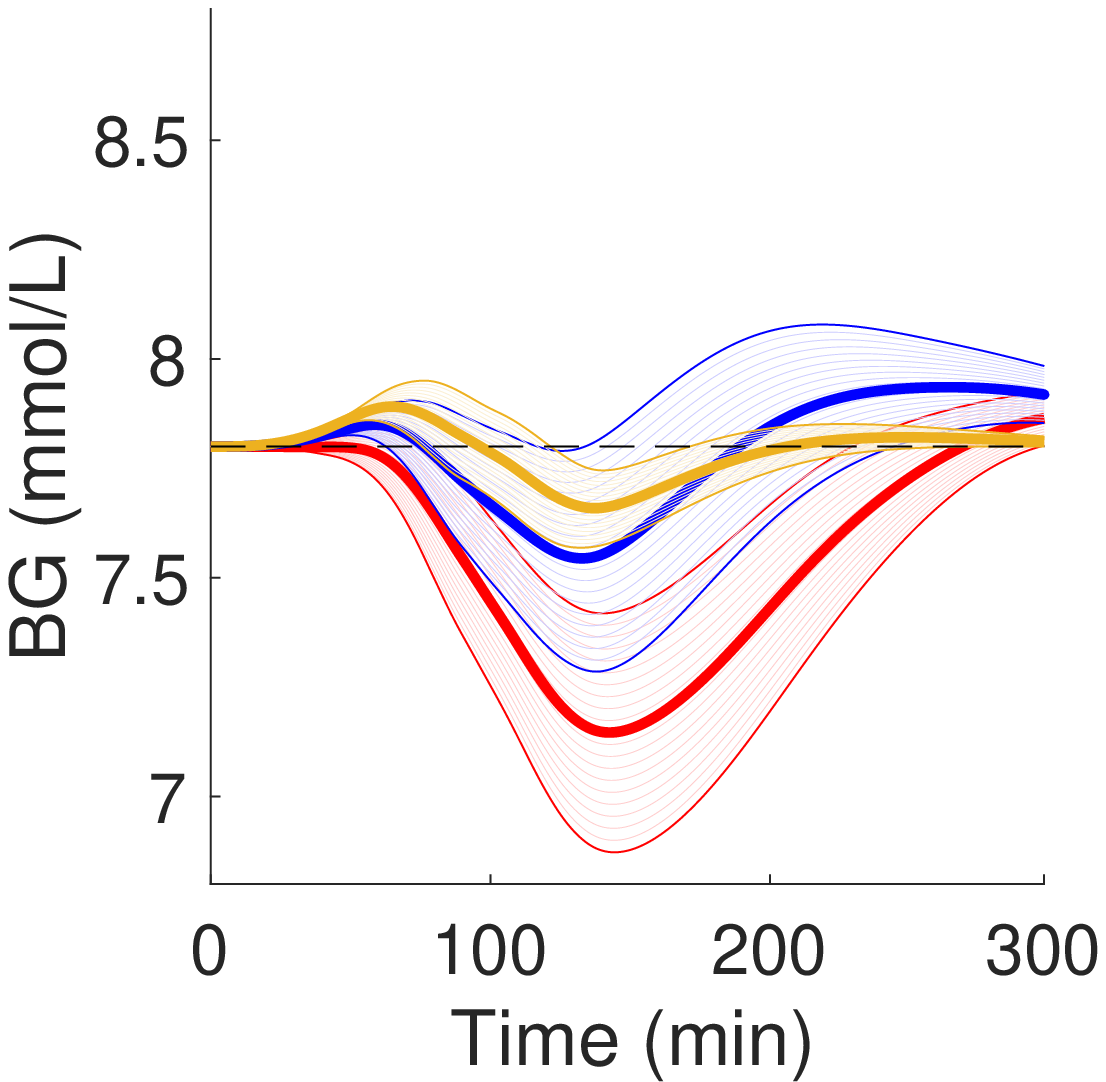}}
\hfill
\subfloat[IIR]{\includegraphics[height=.2\textwidth]{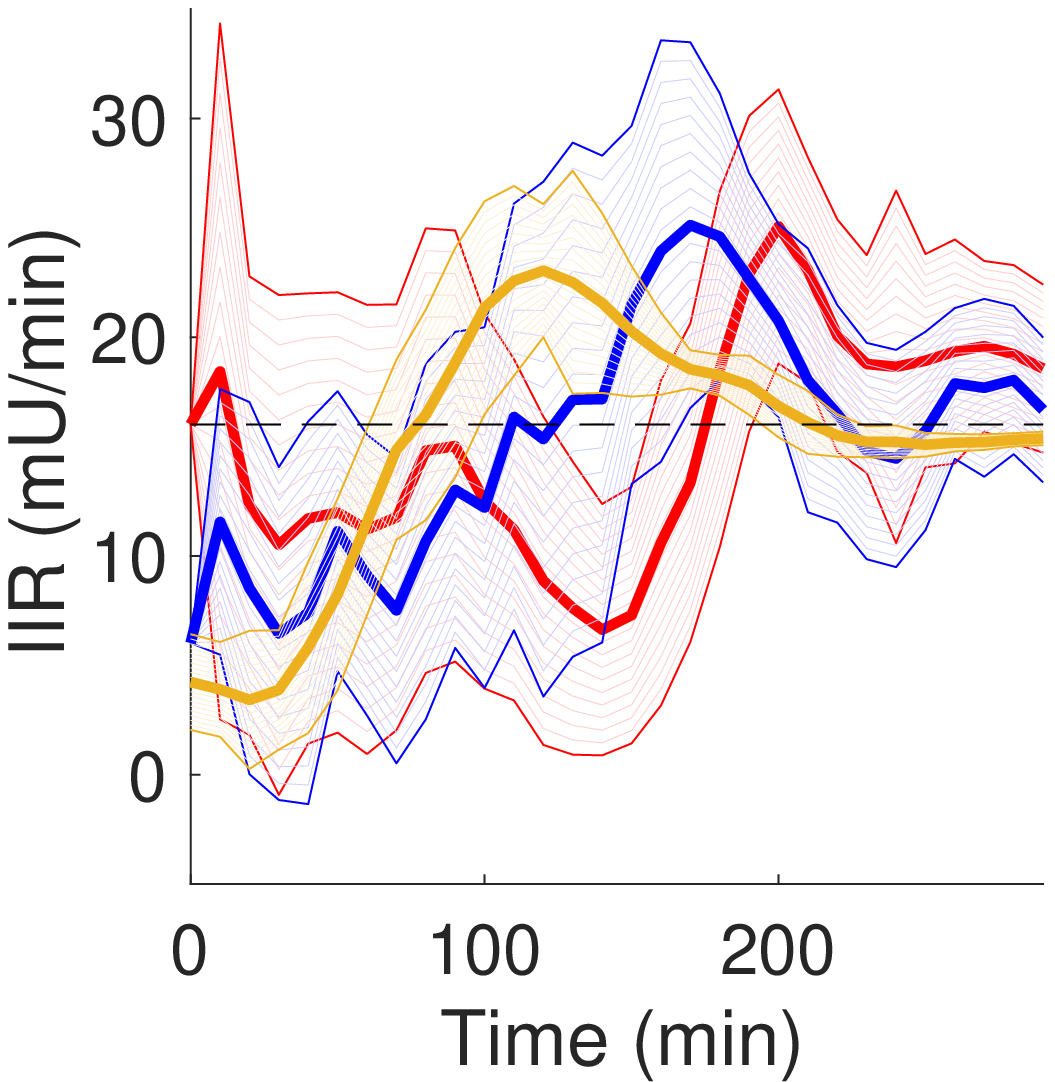}}\\[5pt]
{\footnotesize
\begin{tabular}{r|ccccccccc}
& $t_{\sf < 3.9}$ & $t_{\sf 3.9-11.1}$ & $t_{\sf > 11.1}$ & $BG_{\min}$ & $BG_{\max}$ &$E_{MM}$  &$E_{O2}$ &$E_{BG}$  & $\sum \iota$\\ \hline
Perfect &  0\% &  100\% & 0\% & 7.64 & 7.92 & N.A. & N.A. &  N.A. & -0.29\\
HCL &   0\% &  100\% & 0\% & 7.13& \textbf{7.88} & 0.05 & 6.97\% & 0.42 & -0.26\\ 
Robust &  0\% &  100\% & 0\% & \textbf{7.5} & 7.98 & 0.05 & \textbf{5.04\%} &  0.42 & -0.22\\ \hline
\end{tabular}
}
\caption{Regulation during random exercise (50 repetitions). a) and b) show uncertainty sets and realizations for active muscular mass ($\mathit{MM}$) and oxygen consumption ($\mathit{02}$). Legend is as in Figure \ref{fig:results_table_synth}. \vspace*{-1em}}\label{fig:result_exer}
\end{figure}

\noindent Results, reported in Figure \ref{fig:result_exer}, evidence that both the robust and the HCL controller can maintain BG within very tight ranges, as confirmed by the $BG_{\min}$ and $BG_{\max}$ indicators. BG profiles are almost indistinguishable from the ideal ones (i.e.~those of the perfect controller) and for 100\% of the times within healthy ranges. Note that both controllers correctly reduce the insulin therapy below the basal level to counteract the decrease of BG due to exercise. Hence, the negative values of $\sum \iota$. The main difference is that the robust controller, due to the superior predictive capabilities, is more timely in cutting insulin therapy than the HCL controller, leading to a smaller excursion from the target BG value.

Resalat et al. \cite{resalat2016design} realized a similar scenario to test their dual-hormone MPC (300-minute simulation with a 45-minute exercise at fixed $\mathit{02} = 60$ and $\mathit{MM} = 0.8$). While we use their same plant model, their MPC design is different in two ways: it can regulate both insulin and glucagon (to prevent hypoglycemia) and is not robust, meaning that exercise must be announced in order for the controller to make correct predictions. Despite that, however, their evaluation resulted into some episodes of hypoglycemia and hyperglycemia, while our controller is able to keep BG for 100\% of the time in healthy ranges without meal announcements.

\subsection{One-day experiments using NHANES survey data}\label{sect:NHANES}
\begin{figure}
\centering
\begin{minipage}{.62\textwidth}
\subfloat[$D_G$]{\includegraphics[width=.5\textwidth]{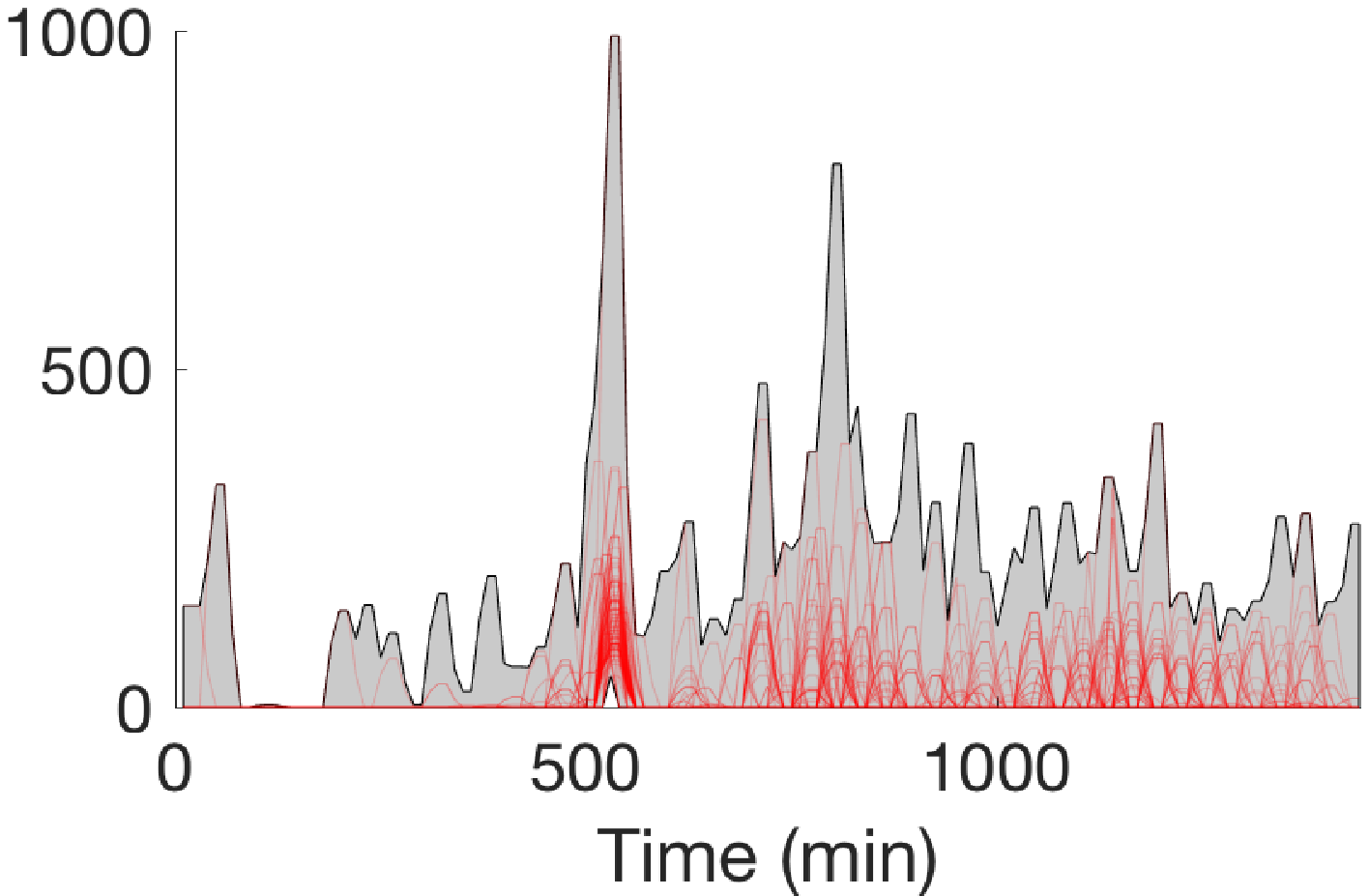}}
\subfloat[BG]{\includegraphics[width=.5\textwidth]{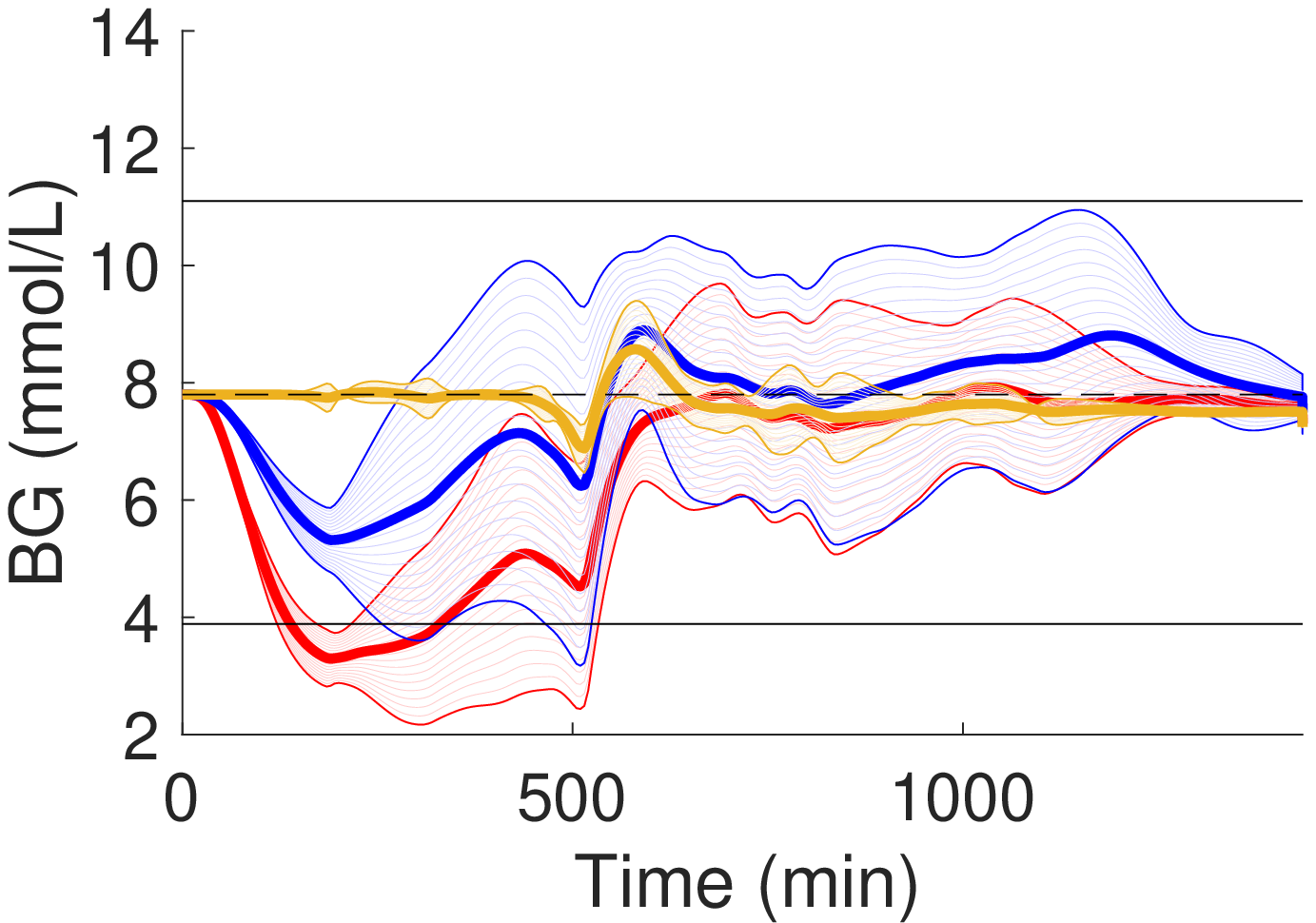}}
\end{minipage}
\hspace*{-.2cm}
\begin{minipage}{.38\textwidth}
\begin{footnotesize}
\begin{tabular}{r|ccc}
& $t_{\sf < 3.9}$ & $t_{\sf 3.9-11.1}$ & $t_{\sf > 11.1}$\\ \hline
Perfect &  0\% &  100\% & 0\% \\
HCL &   18.5\% &  80.97\% & \textbf{0.53}\%\\ 
Robust & \textbf{2.02\%} &  \textbf{93.45\%} & 4.52\%\\ \hline
\end{tabular}
\end{footnotesize}
\end{minipage}
\caption{BG regulation for virtual patient learned from NHANES database (20 repetitions). Legend is as in Figure~\ref{fig:results_table_synth}.}
\label{fig:nhanes_box}
\end{figure}

We test our robust controller with real population data from the CDC's National Health and Nutrition Examination Survey (NHANES) database.\footnote{\url{https://www.cdc.gov/nchs/nhanes/}} We consider the 2013 survey, comprising 8,611 participants, and classify the participants into 10 groups using k-means clustering. 
In this experiment, we selected the cluster 
whose meal patterns are characterized by a CHO-rich breakfast at around 9am, as visible in the uncertainty set of Figure~\ref{fig:nhanes_box}(a).
From this cluster, we extract meal information to parameterize the virtual patient and build the uncertainty sets as explained in Section~\ref{sec:usets} (choosing $\alpha=0.2$ and $\epsilon=0.2$). Due to the poor quality of physical activity data in NHANES, we generated one random exercise event for each patient. 
Details on the other clusters and on extraction and processing of data are given in Appendix~\ref{app:nhanes}. 

Results were obtained with 20 repetitions and are reported in Figure~\ref{fig:nhanes_box}. In this experiment, our robust controller has a close-to-ideal performance, with $>$93\% of time spent in normal BG ranges. It outperforms the HCL controller, which fails to predict the correct BG levels during sleep (time $<$ 500 min), leading to excessive insulin therapy and  to dangerous overnight hypoglycemia. 

\subsection{High carbohydrate intake scenario}\label{sect:high_carbs}

\begin{figure}
\centering
\begin{minipage}{.4\textwidth}
\includegraphics[width=\textwidth]{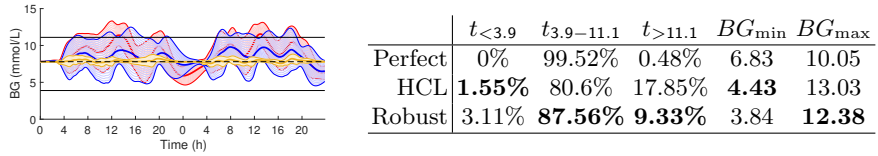}
\end{minipage}
\begin{minipage}{.58\textwidth}
\begin{footnotesize}
\begin{tabular}{r|ccccc}
& $t_{\sf < 3.9}$ & $t_{\sf 3.9-11.1}$ & $t_{\sf > 11.1}$ & $BG_{\min}$ & $BG_{\max}$\\ \hline
Perfect &  0\% &  99.52\% & 0.48\% & 6.83 & 10.05\\
HCL &   \textbf{1.55\%} &  80.6\% & 17.85\% & \textbf{4.43} & 13.03\\ 
Robust &  3.11\% &  \textbf{87.56\%} & \textbf{9.33\%} & 3.84 &\textbf{12.38}\\ 
\hline
\end{tabular}
\end{footnotesize}
\end{minipage}
\caption{BG profile (left) and performance indicators (right) for the high carbohydrate intake scenario (50 repetitions). Legend is as in Fig. \ref{fig:results_table_synth}. \vspace*{-1em}}
\label{fig:high_carb_results}
\end{figure}

We assess the behavior of the controller under irregular meal timing and unusually high CHO intake, following the protocol of \cite{szalay2014linear}, reported in Table \ref{tbl:high_carb_table}. In this protocol, no physical activity is considered. Uncertainty sets were derived following the same construction of the one-meal experiments. Results, obtained with $50$ repetitions, are shown in Figure \ref{fig:high_carb_results}. 

\setlength{\intextsep}{5pt}%
\setlength{\columnsep}{8pt}
\begin{wraptable}{r}{0.5\textwidth}
\centering
\begin{footnotesize}
\begin{tabular}{r|c|c|c}
& Chance of & CHO  & Time of \\
& occurrence &  (g) & day (h)\\ \hline
Breakfast & 100\% & 40-60 & 6:00-10:00\\
Snack 1 & 50\% & 5-25 & 8:00-11:00\\
Lunch & 100\% & 70-110 & 11:00-15:00\\
Snack 2 & 50\% & 5-25 & 15:00-18:00\\
Dinner & 100\% & 55-75 & 18:00-22:00\\
Snack 3 & 50\% & 5-15 & 22:00-00:00\\ \hline
\end{tabular}
\end{footnotesize}
\caption{High carbohydrate intake simulation parameters of \cite{szalay2014linear}. Meals in the plant are sampled uniformly based on the above intervals and probabilities.}\label{tbl:high_carb_table}
\vspace{-0.1em}
\end{wraptable}

Our robust controller resulted in 87.56\% of time within healthy BG ranges, against the 80.6\% of the HCL controller. Despite hypoglycemia amounts to 3.11\% of the total time, it corresponds only to minor episodes, as visible by the standard deviation intervals in the plot and by the average minimum BG ($BG_{\min} = $3.84 mmol/L) that falls only slightly below the hypoglycemic level (3.9 mmol/L).

We also report that our approach outperforms the robust LPV approach of Jacobs et al.~\cite{szalay2014linear}, discussed in the related  work (Section~\ref{sect:related}). With the same plant model and scenario, they obtain $t_{\sf < 3.9}=0$\%, $t_{\sf 3.9-11.1}=83.08$\% and $t_{\sf > 11.1}=16.92$\%, meaning that our robust controller stays $>4\%$ of the time longer in healthy ranges. 
We remark that the results of Jacobs et al.\ are as reported in~\cite{szalay2014linear}, and were not obtained by running their controller on our machine. 


\subsection{Evaluation of state estimator}\label{sect:res_estimation}

We chose an MHE scheme for state estimation (see Section \ref{sec:state_est}) after having evaluated \textit{extended Kalman filters (EKF)} \cite{welch1995introduction}, which are commonly employed for the state estimation of non-linear systems. MHE overcomes some of the typical problems of Kalman filtering, namely, the inability to accurately incorporate state constraints (e.g.~non-negative concentrations); poor use of the nonlinear model \cite{haseltine2005critical}; and estimations that often diverge, or converge to wrong state predictions \cite{van2004sigma,perea2007nonlinearity}. Moreover,``off-the-shelf'' Kalman filters only support zero-mean disturbances (white Gaussian noise), thus preventing the estimation of random meal and exercise episodes. 

\begin{figure}
\centering
\begin{minipage}{.47\textwidth}
\subfloat[$q=0.1521$]{\includegraphics[height=2.5cm]{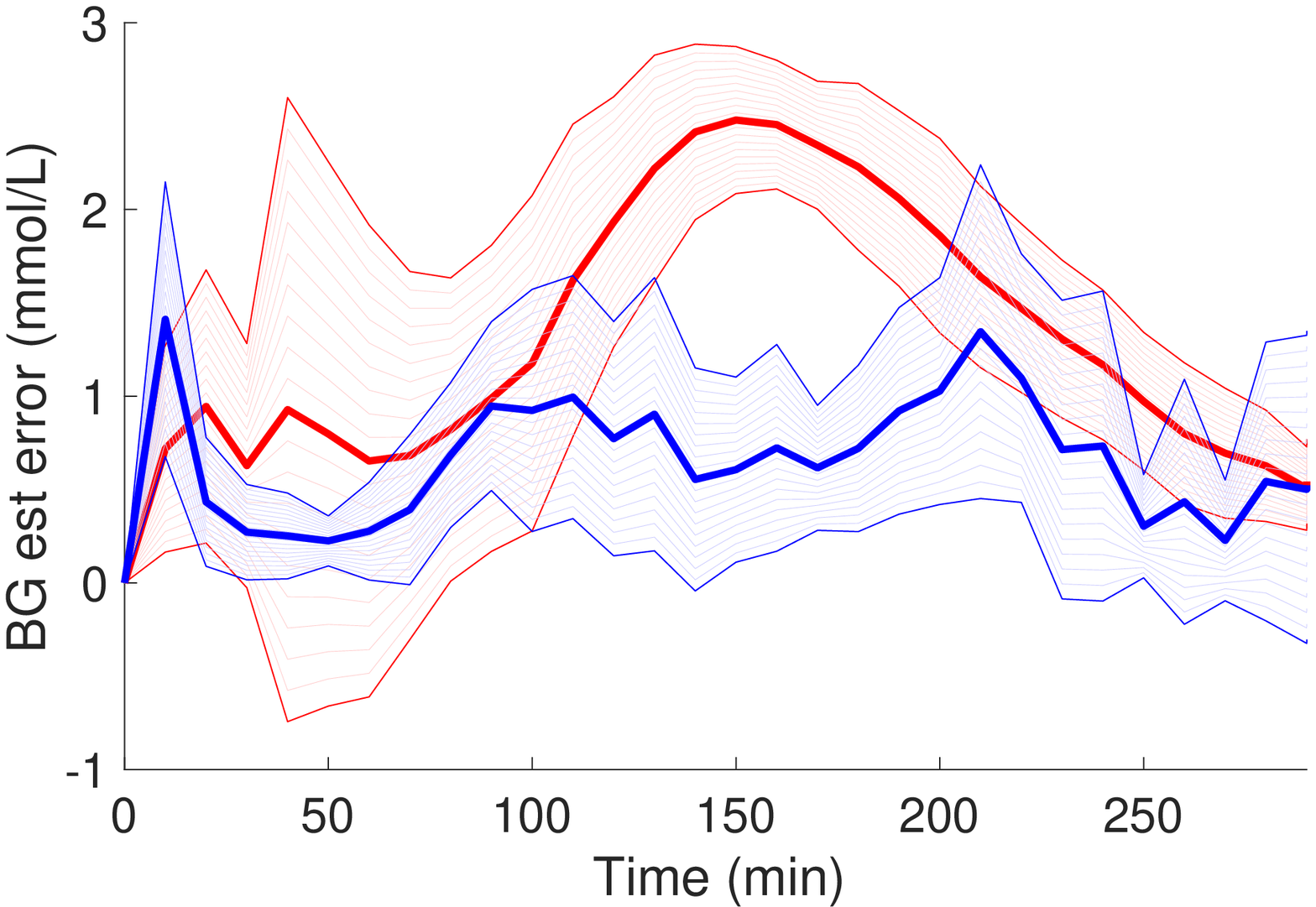}} \\
{\footnotesize
\begin{tabular}{r|ccccc}
& $t_{\sf < 3.9}$ & $t_{\sf 3.9-11.1}$ & $t_{\sf > 11.1}$ &$E_{D_G}$ &$E_{BG}$ \\ \hline
MHE &  0.17\% &  \textbf{96.02\%} & \textbf{3.82\%} & 1.97 & \textbf{0.85}\\
EKF &  \textbf{0\%} & 44.03\% & 55.97\% & N.A. & 1.63\\ \hline
\end{tabular}
}
\end{minipage}\hspace*{.3cm}
\begin{minipage}{.5\textwidth}
\vspace*{-.3cm}
\subfloat[$q=1$]{\includegraphics[height=2.75cm]{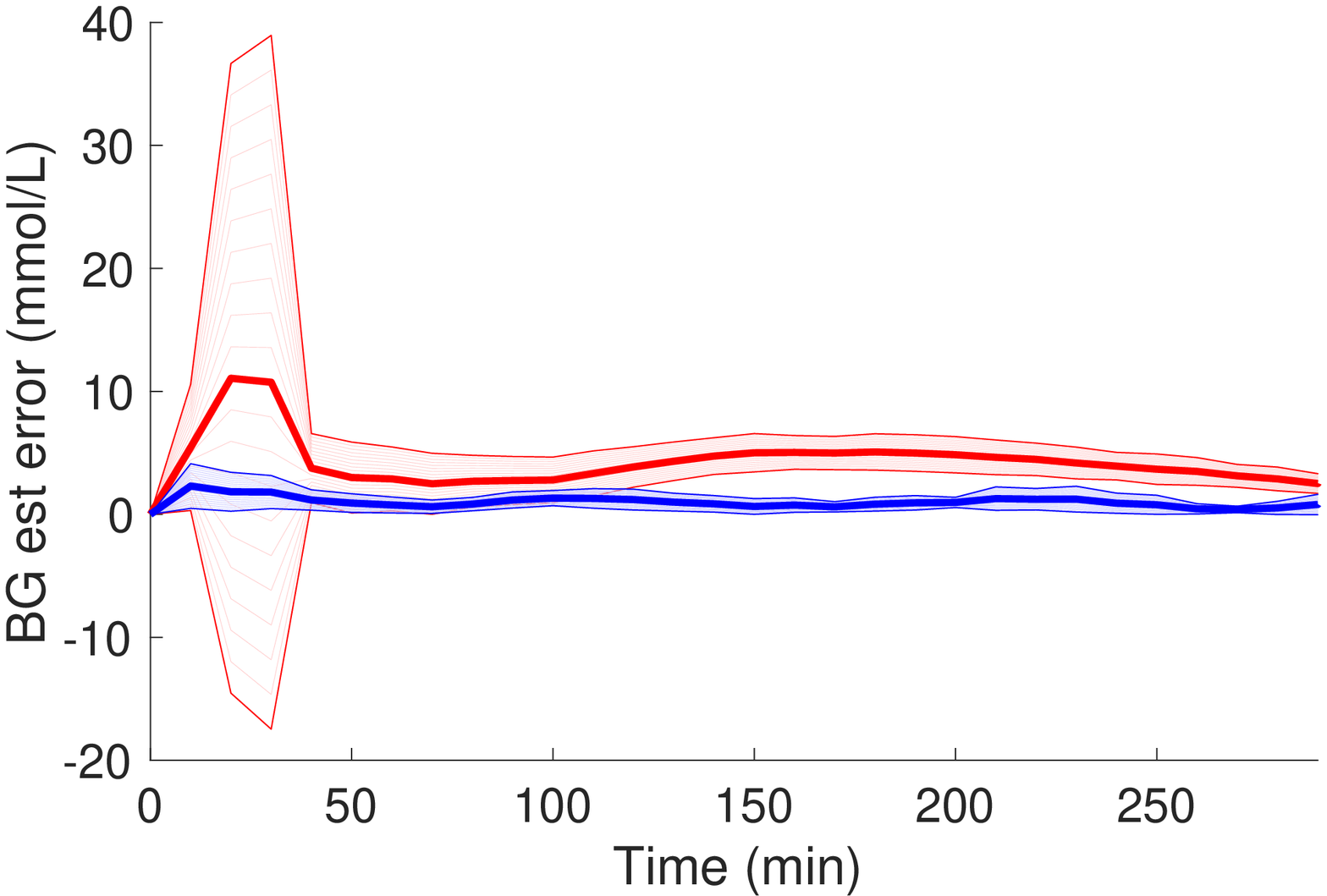}}  \hspace*{-.5cm}
\subfloat{\includegraphics[width=.2\textwidth]{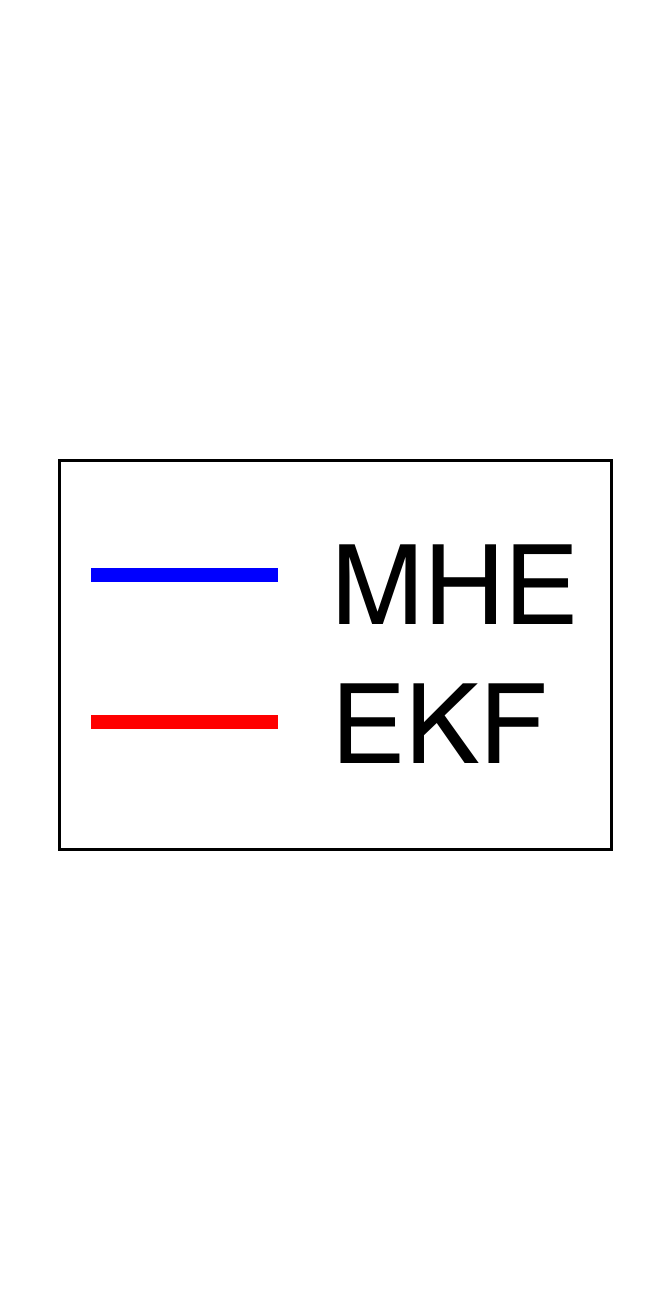}} \\
{\footnotesize
\begin{tabular}{r|ccccc}
& $t_{\sf < 3.9}$ & $t_{\sf 3.9-11.1}$ & $t_{\sf > 11.1}$ &$E_{D_G}$ &$E_{BG}$ \\ \hline
MHE &  \textbf{0\%} & \textbf{94.38\%} & \textbf{5.62\%} & 2 & \textbf{1.15}\\
EKF &  1.32\% & 43.75\% & 54.93\% & N.A. & 4.44\\ \hline
\end{tabular}
}
\end{minipage}
\caption{BG estimation error of Moving Horizon Estimator (MHE) and Extended Kalman Filter (EKF), at different sensing noise variances $q$ (20 repetitions).}\label{fig:estimation_res}
\end{figure}

We compare the state estimation accuracy between our MHE design and an EKF scheme, according to the \textit{meals as expected} scenario (see Section \ref{sect:one_meal_experiments}).  In the EKF, to predict the state estimate at time $t$, $\hat{\bf x}(t)$, we use the model of Section \ref{sec:model} as follows: $\dot{\hat{\mathbf{x}}}(t)= F\left(\hat{\mathbf{x}}(t), \iota^t, \mathrm{E}[\mathbf{u}^t] \right)$, where $\iota^t$ is the (known) insulin input and uncertainty parameters $\mathbf{u}^t$ are replaced with their expected value $\mathrm{E}[\mathbf{u}^t]$\footnote{The real expected value of $\mathbf{u}^t$ is known because here we work with arbitrary distributions.}. 

To evaluate if the estimators are robust with respect to sensing noise, we tested two different variance values for the sensing noise: $q=0.1521$ (default) and $q=1$ (increased noise). As visible in Figure \ref{fig:estimation_res}, the MHE outperforms the EKF, with a consistently lower state estimation error. The imprecise state predictions of the EKF lead to a wrong behavior of the overall closed-loop system, with only $\sim$ 44\% of time spent within normal BG ranges, against $>$ 94\% of the MHE. Unlike the EKF, the MHE is robust to sensing noise, with an average estimation error (column $E_{BG}$) that stays relatively constant from $q=0.1521$ to $q=1$. 
\section{Related Work}\label{sect:related}
Robust control methods are able to minimize the impact of input disturbances on the plant, and thus have the potential to enable fully closed-loop insulin delivery. Earlier approaches \cite{kienitz1993robust,parker2000robust,ruiz2004blood} are based on the theory of $H_{\infty}$ control  \cite{stoorvogel1992hinf}, a technique where the robust controller is synthesized offline as the result of an optimization problem that minimizes the worst-case closed-loop performance of the controlled system. However, $H_{\infty}$ control only supports linear systems, thus requiring linearization of physiological, non-linear gluco-regulatory models, with inevitable loss of accuracy. 

Kovacs et al. \cite{kovacs2011induced,kovacs2013applicability,szalay2014linear} introduce robust linear parameter varying (LPV) control, a technique that consists on deriving a piecewise-linear approximation of the non-linear plant and synthesizing a robust $H_{\infty}$ controller for each linear region, and thus, improves on previous $H_{\infty}$ approaches. 
In Section \ref{sect:high_carbs}, we have compared our robust controller to \cite{szalay2014linear}, showing that our algorithm is able keep glucose levels within normal ranges for a longer time. 

In contrast to the above techniques, our data-driven robust MPC supports not just meal disturbances, but also physical activity, and is based on non-linear optimization, meaning that it does not require to approximate the system dynamics, leading to more precise predictions. Further, MPC is known to be superior for individualized control strategies \cite{magni2009model,wang2010model,de2011modeling}, even though is computationally more demanding than offline techniques like $H_{\infty}$ or LPV control, but still feasible within the update periods typical of the artificial pancreas (5-10 minutes). Finally, our data-driven scheme supports continuous learning of the patient's behavior, thus enabling the synthesis of robust and adaptive insulin therapies. On the other hand, $H_{\infty}$ and LPV controllers are offline and need to be synthesized from scratch in order to adapt to changing patient conditions. 

A simpler strategy employed in a number of AP studies, see e.g.~\cite{laxminarayan2012use,huyett2015design}, is that of PID control, where the control input results from applying tunable gains to the error between the system output and a desired setpoint. Synthesizing these gains to obtain robustness guarantees, however, becomes difficult for systems with nonlinear and probabilistic dynamics.
\section{Conclusions}\label{sec:concl}
Thanks to modern wearable sensing devices, patient-specific data about meals and physical activity is becoming more  readily available, making it possible to offer significantly enhanced personalized medical therapy for type~1 diabetes. Accordingly, we presented a data-driven robust MPC framework for T1D that leverages meal and exercise data to provide enhanced control and state estimation.  Our results show that learning a patient's behavior from data is key to achieving fully closed-loop therapy that does not require meal and exercise announcements.

\paragraph{Acknowledgments} Research supported in part by %
AFOSR Grant FA9550-14-1-0261 
and NSF Grants
IIS-1447549, 
CNS-1446832, 
CNS-1445770, 
CNS-1445770, 
CNS-1553273, CNS-1536086, and IIS-1460370.

\bibliographystyle{abbrv}
\bibliography{pancreas}

\appendix
\section{Gluco-regulatory ODE model}\label{app:full_ODE}
We describe the details of the ODE system used in our controller design, which consists of the following subsystems:

\paragraph{Glucose kinetics} describes the glucose masses in the accessible (where BG measurements are made) and non-accessible compartments, respectively through variables $Q_1$ and $Q_2$ (mmol) as follows:

{\vspace*{-8pt}\footnotesize
\begin{align}\label{eq:glu_kin}
\dot{Q}_1(t) = & -F_{01c}  -x_1 \cdot Q_1(t) + k_{12}\cdot Q_2(t) - F_R + U_g(t) + EGP_0\cdot (1 - x_3(t)) \\
\dot{Q}_2(t) = & x_1(t)\cdot Q_1(t) - k_{12}\cdot Q_2(t) -x_2(t)\cdot Q_2(t) \nonumber
\end{align}}\noindent
where $F_{01c}$ and $F_R$ (mmol min$^{-1}$) are the corrected non-insulin mediated glucose uptake and renal glucose clearance, respectively, derived as per~\cite{hovorka2004nonlinear}; $x_1,x_2,x_3$ describe the effect of insulin on glucose (see the \textit{insulin dynamics} subsystem); $U_g$ is the gut absorption rate (see the \textit{gut absorption} subsystem); and $EGP_0$ (mmol min$^{-1}$) is the glucose production at a theoretical zero-insulin concentration. BG concentration, $G$ (mmol L$^{-1}$), is derived from $Q_1$ as $G(t) = Q_1(t)/V_G$, where $V_G$ is the glucose distribution volume. In our robust MPC controller (Section \ref{sec:rob_mpc}), $G$ is the main state variable that we want to control.

\paragraph{Interstitial glucose} subcutaneous glucose concentration $C$ (mmol/L) detected by the CGM sensor has a delayed response w.r.t.\ the blood concentration $G$, and is given by:

{\vspace*{-5pt}\footnotesize
\begin{equation}
\dot{C}(t) = k_{a\_int} \cdot (G(t) - C(t))
\end{equation}}\noindent
Therefore, the measurement function $h$ of Eq.~\ref{eq:output} maps the state vector at time $t$ to $C(t)$.

\paragraph{Gut absorption} this subsystem uses a chain of two compartments, $G_1$ and $G_2$ (mmol), to model the absorption dynamics of ingested food (given by the uncertainty parameter $D_G^t$) \cite{wilinska2010simulation}:

{\vspace*{-8pt}\footnotesize
\begin{equation}
\dot{G}_1(t) = -G_1(t)/T_{\max} + A_g\cdot D_G^t, \quad \dot{G}_2(t) = (G_1(t) - G_2(t))/T_{\max}
\end{equation}}\noindent
where $A_g$ (unitless) is the CHO bio-availability, and $T_{\max}$ (mins) is the time of maximum appearance rate of glucose, computed as per \cite{wilinska2010simulation}. The gut absorption rate $U_g(t) = G_2(t)/T_{\max}$ (mmol min$^{-1}$) characterizes the flow into the plasma glucose compartment $Q_1$ (see Eq.~\ref{eq:glu_kin}). 

\paragraph{Insulin kinetics} models the absorption of the fast-acting insulin $\iota^t$ (i.e.\ our control input, in mU min$^{-1}$) and its transport through compartments $Q_{1a}$, $Q_{1b}$, $Q_{2i}$ and $Q_{3}$ (in mU) \cite{wilinska2005insulin}:

{\vspace*{-8pt}\footnotesize
\begin{align}
\dot{Q}_{1a}(t) = & K\cdot \iota^t - k_{ia1}\cdot Q_{1a}(t) - \frac{V_{\max,LD} \cdot Q_{1a}(t)}{k_{m,LD} + Q_{1a}(t)}\\
\dot{Q}_{1b}(t) = & (1-K)\cdot \iota^t - k_{ia2}\cdot Q_{1b}(t) - \frac{V_{\max,LD} \cdot Q_{1b}(t)}{k_{m,LD} + Q_{1b}(t)} \nonumber\\
\dot{Q}_{2i}(t) = & k_{ia1}\cdot Q_{1a}(t)  -k_{ia1}\cdot Q_{2i}(t) \nonumber\\
\dot{Q}_3(t) = & k_{ia1}\cdot Q_{2i}(t) + k_{ia2}\cdot Q_{1b}(t) -k_{e}\cdot Q_{3}(t) \nonumber
\end{align}}\noindent
This model assumes a slow insulin absorption pathway consisting of compartments $Q_{1a}$ (subcutaneous insulin mass) and $Q_{2i}$ (non-accessible insulin), and a fast pathway that includes compartment $Q_{1b}$ (subcutaneous). $K$ represents the proportion in which the input insulin $\iota^t$ is distributed into the two pathways. $Q_3$ is the plasma insulin mass. The plasma insulin concentration $I$ (mU L$^{-1}$) is derived as $I(t) = Q_3(t)/V_I$, where $V_I$ is the insulin distribution volume. $V_{\max,LD}$ (mU min$^{-1}$) and $k_{m,LD}$ (mU) are the Michaelis-Menten constants characterizing local insulin degradation. 

\paragraph{Insulin dynamics} defines the effects of insulin on glucose distribution with variable $x_1$ (min$^{-1}$), on glucose disposal with $x_2$ (min$^{-1}$), and on the endogenous glucose production $x_3$ (unitless):

{\vspace*{-8pt}\footnotesize
\begin{align}
\dot{x}_1(t) = &  k_{a1} \cdot \left( - x_1(t) + M_{PGU}(t)\cdot M_{PIU}(t) \cdot S_{IT} \cdot I(t) \right)\\
\dot{x}_2(t) = &  k_{a2}\cdot \left( - x_2(t) + M_{PGU}(t)\cdot M_{PIU}(t) \cdot S_{ID} \cdot I(t) \right) \nonumber \\
\dot{x}_3(t) = & k_{a3}\cdot \left( - x_3(t) + M_{HGP}(t)\cdot S_{IE} \cdot I(t) \right) \nonumber
\end{align}}\noindent
where $M_{PGU}$, $M_{PIU}$ and $M_{HGP}$ (unitless) are factors depending on the patient's physical activity (described below). 

\paragraph{Physical activity} consists of two state variables: the glucose uptake due to active muscular tissue $\mathit{UA}$ (mg min$^{-1}$), and the experienced activity level, which is captured by the percentage of maximum oxygen consumption $\mathit{O2}_m$ (unitless): 

{\vspace*{-8pt}\footnotesize
\begin{equation}\label{eq:phys_act}
\dot{\mathit{UA}}(t) = k_{\mathit{UA}} \cdot (\overline{\mathit{UA}}(t) - \mathit{UA}(t)), \quad 
\dot{\mathit{O2}}_m(t) = k_{O2} \cdot (\mathit{O2}_m(t) - \mathit{O2}^t)
\end{equation}}\noindent
where $\mathit{O2}^t$ is the input uncertainty parameter describing the target workload, and $\overline{\mathit{UA}}(t) = f\left(\mathit{O2}_m(t)\right)$ is the steady-state value of $\mathit{UA}$ which is computed as a function of $\mathit{O2}_m$, where $f$ is estimated in~\cite{lenart2002modeling} using quadratic regression. 

The effects of exercise on peripheral glucose uptake ($M_{PGU}$), on peripheral insulin uptake ($M_{PGU}$), and on hepatic glucose production ($M_{HPG}$) are affected by $\mathit{UA}$ and $\mathit{O2}_m$ as follows:

{\vspace*{-8pt}\footnotesize
\begin{align}
M_{PGU}(t) = & 1 + k_{PGU} \cdot \mathit{UA}(t)\cdot \mathit{MM}^t \\
M_{PIU}(t) = & 1 + k_{PIU} \cdot \mathit{MM}^t \nonumber \\  
M_{HPG}(t) = & 1 + k_{HPG} \cdot \mathit{UA}(t)\cdot \mathit{MM}^t \nonumber
\end{align}}\noindent
where $\mathit{MM}^t$ is the uncertainty parameter for the active muscular mass. 

\begin{sidewaystable}
\begin{footnotesize}
\begin{tabular}{llllp{13cm}}
& \textbf{Value} & \textbf{Unit} & \textbf{Ref} & \textbf{Description}\\ \hline
\multicolumn{5}{c}{\textbf{Glucose Kinetics}}\\ \hline
$F_{01}$ & $0.0104\cdot BW$ & [mmol min$^{-1}$] & \cite{jacobs2015incorporating} & Non-insulin mediated glucose uptake\\ 
$F_{01,thr}$ & $4.5$ & [mmol L$^{-1}$] & \cite{hovorka2004nonlinear} & Non-insulin mediated glucose uptake threshold\\ 
$EGP_0$ & $0.0158\cdot BW$ & [mmol min$^{-1}$] & \cite{jacobs2015incorporating} & Endogenous glucose production extrapolated to zero insulin concentration\\ 
$k_{12}$ & $0.0793$ & [min$^{-1}$] & \cite{jacobs2015incorporating} & Inflow rate from non-accessible compartment\\ 
$V_G$ & $0.1797\cdot BW$ & [L] & \cite{jacobs2015incorporating} & Glucose distribution volume\\ 
$R_{thr}$ & 9 & [mmol L$^{-1}$] & \cite{hovorka2004nonlinear} & Renal clearance threshold\\ 
$R_{cl}$ & 0.003 & [min$^{-1}$] & \cite{hovorka2004nonlinear} & Renal clearance rate\\ 
\hline
\multicolumn{5}{c}{\textbf{Insulin Kinetics}}\\ \hline
$K$ & 0.7958 & unitless & \cite{jacobs2015incorporating} & Proportion of insulin in slow compartment\\ 
$k_{ia1}$ & 0.0113 & [min$^{-1}$] & \cite{jacobs2015incorporating} & Rate constant - slow insulin compartment\\ 
$k_{ia2}$ & 0.0197 & [min$^{-1}$] & \cite{jacobs2015incorporating} & Rate constant - fast insulin compartment\\ 
$k_{e}$ & 0.1735 & [min$^{-1}$] & \cite{jacobs2015incorporating} & Plasma insulin elimination rate\\ 
$V_{\max,LD}$ & 2.9639 & [mU min$^{-1}$] & \cite{jacobs2015incorporating} & Insulin max velocity of local degradation / saturation level\\ 
$k_{m,LD}$ & 47.5305 & [mU] & \cite{jacobs2015incorporating} & Michaelis constant - insulin\\ 
\hline
\multicolumn{5}{c}{\textbf{Insulin Dynamics}}\\ \hline
$k_{a1}$ & 0.007 & [min$^{-1}$] & \cite{jacobs2015incorporating} & Rate of effect of remote insulin on glucose distribution/transport\\ 
$k_{a2}$ & 0.0331 & [min$^{-1}$] & \cite{jacobs2015incorporating} & Rate of effect of remote insulin on glucose disposal\\ 
$k_{a3}$ & 0.0308 & [min$^{-1}$] & \cite{jacobs2015incorporating} & Rate of effect of remote insulin on EGP suppression\\ 
$S_{IT}$ & 0.0046 & [min$^{-1}$ mU$^{-1}$ L] & \cite{jacobs2015incorporating} & Insulin sensitivity on glucose distribution/transport\\ 
$S_{ID}$ & 0.0006 & [min$^{-1}$ mU$^{-1}$ L] & \cite{jacobs2015incorporating} & Insulin sensitivity on glucose disposal\\ 
$S_{IE}$ & 0.0384 & [mU$^{-1}$ L] & \cite{jacobs2015incorporating} & Insulin sensitivity on EGP suppression\\ 
$V_I$ & $0.1443\cdot BW$ & [L] & \cite{jacobs2015incorporating} & Insulin distribution volume\\ 
\hline

\multicolumn{5}{c}{\textbf{Gut absorption}}\\ \hline
$A_g$ & 0.8121 & [unitless] & \cite{jacobs2015incorporating} &  Carbohydrate bio-availability of the ingested food (Proportion of absorbed carbs)\\
$U^{\top}_g$ & 0.0275$\cdot BW$ & [mmol min$^{-1}$] & \cite{wilinska2010simulation} & Maximum glucose flux from the gut\\
$T^{\bot}_{\max}$ & 48.8385 & [mmol min$^{-1}$] & \cite{jacobs2015incorporating} & Lower bound for time-of-maximum appearance rate of glucose in the accessible compartment\\ \hline

\multicolumn{5}{c}{\textbf{Interstitial (sensor) glucose}}\\ \hline
$k_{a\_int}$ & 0.025 & [min$^{-1}$] & \cite{wilinska2010simulation} & Peripheral glucose uptake factor\\ \hline

\multicolumn{5}{c}{\textbf{Exercise}}\\ \hline
$k_{UA}$ & $\frac{1}{30}$ & [min$^{-1}$] & \cite{jacobs2015incorporating} & Rate affecting the dynamics of peripheral glucose uptake during exercise\\ 
$k_{O2}$ & $\frac{5}{3}$ & [min$^{-1}$] & \cite{lenart2002modeling} & Rate affecting the time needed to reach the target exercise level\\ 
$k_{MPGU}$ & 35 & [mg min$^{-1}$] & \cite{jacobs2015incorporating} & Basal peripheral glucose uptake\\
$k_{HPG}$ & 155 & [mg min$^{-1}$] & \cite{jacobs2015incorporating} & Basal hepatic glucose production\\
$k_{PIU}$ & 2.4 & unitless & \cite{jacobs2015incorporating} & Peripheral insulin uptake factor\\
$a_{UA-O2}$ & 0.006 & [mg min$^{-1}$] & \cite{jacobs2015incorporating} & Quadratic coefficient in $UA$-$O2_{m}$ law\\
$b_{UA-O2}$ & 1.2264 & [mg min$^{-1}$] & \cite{jacobs2015incorporating} & Linear coefficient in $UA$-$O2_{m}$ law\\
$c_{UA-O2}$ & -10.1958 & [mg min$^{-1}$] & \cite{jacobs2015incorporating} & Constant coefficient in $UA$-$O2_{m}$ law\\
\hline
\end{tabular}
\end{footnotesize}
\caption{Model parameters. $BW$ (kg) is the body weight of the virtual patient. For our experiments, we set $BW = 75$ kg.}\label{tbl:model_params}
\end{sidewaystable}

\section{Construction of uncertainty sets from data}
We will describe the assumptions and the details of the construction for the box type uncertainty sets. We follow the notations similar in \cite{bertsimas2013data}. $\mathbf{u} \in \mathbb{R}^d$ denotes the random uncertainty vector and $u_i$ denotes its components. $\mathbb{P}^*$ refers to the true and unobserved probability measure for $\mathbf{u}$. The set of sample data points $\mathcal{S} = \{ \hat{\mathbf{u}}^1,\ldots,\hat{\mathbf{u}}^S \}$ is constructed by drawing i.i.d. $S = |\mathcal{S}|$ times from $\mathbb{P}^*(\mathbf{u})$. Here, we do not need to assume the marginal distributions of $\mathbb{P}^*$ to be independent. That is consistent to the observation that the elements in uncertainty parameter $\mathbf{u}$ for artificial pancreas controller is correlated. Moreover, the box type is designed to be suitable for the case when the sample data contains many missing values or we are only able to collect samples asynchronously.


\subsubsection{Box type}
The univariate hypothesis test for the $1-\epsilon/d$ quantile in David and Nagaraja is extended to the multivariate case in \cite{bertsimas2013data}. Given $\bar{q}_{i,0}, \underline{q}_{i,0} \in \mathbb{R}, \forall i=1,\ldots,d$,
\begin{align*}
H_0 :& \inf\{v:\mathbb{P}(u_i\leq v)\geq 1-\epsilon/d\} \geq \bar{q}_{i,0} \text{ and } \\
&\inf\{v:\mathbb{P}(-u_i\leq v)\geq 1-\epsilon/d\} \geq \underline{q}_{i,0}\forall i=1,\ldots,d
\end{align*}
Assume that we have $S$ random samples. The index $s$ is defined as
$$ s = \min \left\{ k \in \mathbb{N}:\sum^{S}_{j=k} \binom{S}{j} (\frac{\epsilon}{d})^{S-j} (1-\frac{\epsilon}{d})^j \leq \frac{\alpha_h}{2d} \right\}. $$
For each component $u_i$ of $\mathbf{u}$ and we re-order them in an increasing order $u_i^{(1)},  u_i^{(2)}, \ldots ,  u_i^{(S)}$. So the test is rejected at level $\alpha_h$ if 
$$\hat{u}_i^{(s)} < \bar{q}_i \text{ or } -\hat{u}_i^{(S-s+1)} < \underline{q}_i$$
by union bound. By this construction, the Theorem 7 in \cite{bertsimas2013data} shows that
if $s$ is defined by the above equation satisfies $S-s+1<s$, then, with probability at least $1-\alpha_h$ over the sample, the set
$$ \mathcal{U}_{\epsilon}^{M} = \left\{\mathbf{u}\in\mathbb{R}^d: \hat{u}_i^{(S-s+1)} \leq u_i \leq \hat{u}_i^{(s)} \forall i = 1,\ldots,d \right\} $$
implies a probabilistic guarantee for $\mathbb{P}^*$ at level $\epsilon$.

\section{Building uncertain sets from probabilistic models }\label{sec:usets_short} 
We show how to build tight time-varying uncertainty sets when the uncertain meal and exercise episodes are uniformly or normally distributed. Importantly, such distributions can also be derived from sample data through the bootstrapping method \cite{efron1994introduction}, as done in the robust taxi dispatch approach of \cite{miao2015robust}. 

For each meal, we assume that the start time, $t_s$, and the total amount of ingested carbohydrates, $\mathit{CHO}$, are uncertain. Meal duration ($d$) is fixed, during which carbohydrate ingestion happens at a constant rate. Similarly, each exercise episode has uncertain start time $t_s$, percentage of muscular mass $\mathit{MM}$, percentage of maximum oxygen consumption $\mathit{O2}$ and duration $d$. 

According to which distribution the meal or exercise event is sampled from, we derive the lower and upper bound of the corresponding uncertainty parameters. The intuition is that when at time $t$ a random variable, say $\mathit{O2}$, is uniformly distributed in the interval $[\mathit{O2}^\bot,\mathit{O2}^\top]$, written as $\mathit{O2} \sim \text{Unif}(\mathit{O2}^\bot,\mathit{O2}^\top)$, then the lower and upper bound is $[\mathit{O2}^\bot,\mathit{O2}^\top]$. Note that such a defined uncertain set covers all possible realizations of the random variable. Instead, when $\mathit{O2}$ is normally distributed with mean $\mu_{\mathit{O2}}$ and standard deviation $\sigma_{\mathit{O2}}$, written as $\mathit{O2} \sim \mathcal{N}(\mu_{\mathit{O2}}, \sigma_{\mathit{O2}})$, we set the bounds to be $[\mu_{\mathit{O2}} - k\cdot \sigma_{\mathit{O2}}, \mu_{\mathit{O2}} + k\cdot \sigma_{\mathit{O2}}]$, with $k>0$. Since the normal distribution has unbounded support, we cannot cover all possible realizations with an interval. In our experiments, we select $k=3$, which covers $\approx 99.74\%$ of all possible values.

Besides the lower and upper bounds, the uncertain sets $\mathcal{U}^t$ are also conditioned by the range of the (uncertain) start time $[t_s^\bot, t_s^\top]$ and the range of duration $[d^\bot, d^\top]$ for each meal/exercise episode. (The duration of meal is fixed so the range is only a single value $d$). Table \ref{tbl:usets} illustrates the rule for the construction of uncertainty set at time $t$.

\begin{table}
\centering
\begin{footnotesize}
\begin{tabular}{c | l }
\hline
$\mathcal{U}^t[D_G] $ 	& ${D_G^t}^\bot = \mathit{CHO}^\bot/d$ if $t_s^\top \leq i \leq  t_s^\bot + d$, $0$ o/w 
\\
$= [{D_G^t}^\bot, {D_G^t}^\top]$ & ${D_G^t}^\top = \mathit{CHO}^\top/d$ if $t_s^\bot \leq i \leq  t_s^\top + d$, $0$ o/w 
\\ \hline

$\mathcal{U}^t[\mathit{MM}] $ 	& ${\mathit{MM}^t}^\bot = \mathit{MM}^\bot$ if $t_s^\top \leq i \leq  t_s^\bot + d^\bot$, $0$ o/w \\
$= [{\mathit{MM}^t}^\bot, {\mathit{MM}^t}^\top]$	& ${\mathit{MM}^t}^\top = \mathit{MM}^\top$ if $t_s^\bot \leq i \leq  t_s^\top + d^\top$, $0$ o/w \\ \hline

$\mathcal{U}^t[\mathit{O2}] $ 	& ${\mathit{O2}^t}^\bot = \mathit{O2}^\bot$ if $t_s^\top \leq i \leq  t_s^\bot + d^\bot$, $\mathit{O2}_0$ o/w \\
$= [{\mathit{O2}^t}^\bot, {\mathit{O2}^t}^\top]$	& ${\mathit{O2}^t}^\top = \mathit{O2}^\top$ if $t_s^\bot \leq i \leq  t_s^\top + d^\top$, $\mathit{O2}_0$ o/w \\ \hline
\end{tabular}
\end{footnotesize}
\caption{Uncertain sets at time $t$ for CHO ingestion rate $D_G$, active muscular mass $\mathit{MM}$ and oxygen consumption $\mathit{O2}$. $\mathit{O2}_0=8$ is the basal oxygen consumption at rest.} \label{tbl:usets}
\end{table}


\begin{table}
\centering
\begin{footnotesize}
\begin{tabular}{r|cccccc}
& $t_{\sf < 3.9}$ & $t_{\sf 3.9-11.1}$ & $t_{\sf > 11.1}$ & $BG_{\min}$ & $BG_{\max}$  & $\sum \iota$\\ \hline
Scenario 1, perfect &  0\% &  99.69\% & 0.31\% & 7.15 & 9.91 & 4.38\\
Scenario 1, HCL &  1.6\% & 69.4\% & 29\% & \textbf{5.61}& 12.85 &  8.19\\ 
Scenario 1, robust &  \textbf{0.51\%} &\textbf{ 97.7\%} & \textbf{1.79\%} & 5.57 & \textbf{9.96} & 6.23\\ \hline
Scenario 2, perfect &  0\% &  100\% & 0\% & 7.03 & 8.84 &  4.67\\
Scenario 2, HCL &  1.03\% & 81.51\% & 17.45\% & \textbf{5.75}& 11.32 & 6.31\\ 
Scenario 2, robust & \textbf{0.28}\% & \textbf{84.19}\% & \textbf{15.53}\% & 5.16 & \textbf{10.94} & 5.82\\ \hline
Scenario 3, perfect &  0\% &  100\% & 0\% & 7.22 & 9.3 & 5.06\\
Scenario 3, HCL &  \textbf{0\%} & 67.25\% & 32.75\% & \textbf{7.19} & 13.34 & 5.05\\ 
Scenario 3, robust &  0.79\% & \textbf{99.03\%} & \textbf{0.18\%} & 5.09 & \textbf{8.77} &  5.64\\ \hline
\end{tabular}
\end{footnotesize}
\caption{Complete statistics for the one-meal experiments of Section \ref{sect:one_meal_experiments}.}
\label{tbl:full_stat_one_meal}
\end{table}

\section{Asymmetric costs}\label{sect:asymm_costs}
We evaluate glucose regulation under different asymmetric costs. By choosing $\gamma>1$ in the controller (see Equation \ref{eq:cost1}), predicted BG trajectories below the target BG level are penalized more than those above the target. As discussed in \cite{gondhalekar2016periodic}, this strategy contributes to reducing hypoglycemic episodes and is substantiated by the fact that hypoglycemia leads to more severe consequences than those of (temporary) hyperglycemia. 

\setlength{\intextsep}{5pt}%
\setlength{\columnsep}{8pt}
\begin{wraptable}{r}{0.5\textwidth}
\centering
\begin{footnotesize}
\begin{tabular}{r|ccccc}
& $t_{\sf < 3.9}$ & $t_{\sf 3.9-11.1}$ & $t_{\sf > 11.1}$ & $BG_{\min}$ & $BG_{\max}$\\ \hline
$\gamma=1$ & 1.5\% & 85.35\% & 13.15\% & 5.17 & 11.05\\
$\gamma=2$ & 0\% & 80.13\% & 18.87\% & 5.4 & 11.38 \\
$\gamma=4$ &  0\% & 76.8\% & 23.2\% & 5.6 & 11.63\\ \hline
\end{tabular}
\end{footnotesize}
\caption{Indicators for different asymmetric cost strategies.}\label{tbl:asymmetric}
\vspace{-0.5em}
\end{wraptable}

We tested the robust controller with $\gamma=1, 2, 4$ (symmetric, 2x, and 4x penalty, respectively). Simulations were conducted according to the \textit{outliers} scenario (see Section \ref{sect:one_meal_experiments}), which typically generates hypoglycemic episodes and thus, is an ideal testbed for tuning $\gamma$.  
Table \ref{tbl:asymmetric} reports the performance indicators obtained with 20 repetitions for each value of $\gamma$. While for $\gamma=1$ (symmetric cost) we record some minor hypoglycemic episodes, hypoglycemia is totally avoided for $\gamma=2,4$.  Between these two values, we chose $\gamma=2$ since it yields smaller hyperglycemia. The indicators for average BG peaks and lows confirm that glucose levels increase with $\gamma$.


\section{Data extraction from NHANES database}\label{app:nhanes}
Below, we describe how we extracted meal data from the CDC's NHANES database and generated the corresponding uncertainty sets.
\begin{itemize}
\item We retrieve meal information from the dietary interview, where each participant reports the timings, types and amounts of each meal during a typical day. 
\item Through a moving average filter, we transform the meal events of each participant into a one-day trajectory describing the CHO intake rate, so that it can be mapped into the uncertainty parameter $D_G$. 
\item Note that a building a single uncertainty set built from the whole database would result in a overly-conservative sets that allows for essentially unrestricted random behaviors. To avoid this, we classify the database based on the above CHO trajectories into 10 groups  using k-means clustering, and select a cluster consisting of 274 people. 
\item Such data is then used to construct the uncertainty sets, as described in Section \ref{sec:usets}, and to parameterize the virtual patient, where random meal uncertainties are sampled from the set of participants. 
\item Due to the lack of good quality data for physical activity in the NHANES database, we generated synthetic exercise data (1 random one-hour exercise episode for each patient) as follows:
\begin{itemize}
\item draw uniformly a random start time for exercise between 9am and 6pm
\item set CHO intake rate to zero for the corresponding time window, since it is unlikely if not impossible that eating and exercise happen at the same time
\item uniformly sample among light, moderate and intense exercise
\item depending on the above outcome, sample oxygen consumption and active muscular mass according to the below predefined ranges: 
\begin{itemize}
\item light: $\mathit{MM} = \mathrm{unif}(0.1,0.25), \mathit{O2} = \mathrm{unif}(15,45)$; 
\item moderate: $\mathit{MM} = \mathrm{unif}(0.2,0.35), \mathit{O2} = \mathrm{unif}(45,75)$;
\item intense: $\mathit{MM} = \mathrm{unif}(0.3,0.5), \mathit{O2} = \mathrm{unif}(75,100)$.
\end{itemize}
\end{itemize}
\end{itemize}

\begin{figure}
\centering
\newcommand{\relheight}{.27}
\subfloat[Cluster \# 2: 368 people]{\includegraphics[height=\relheight\textwidth]{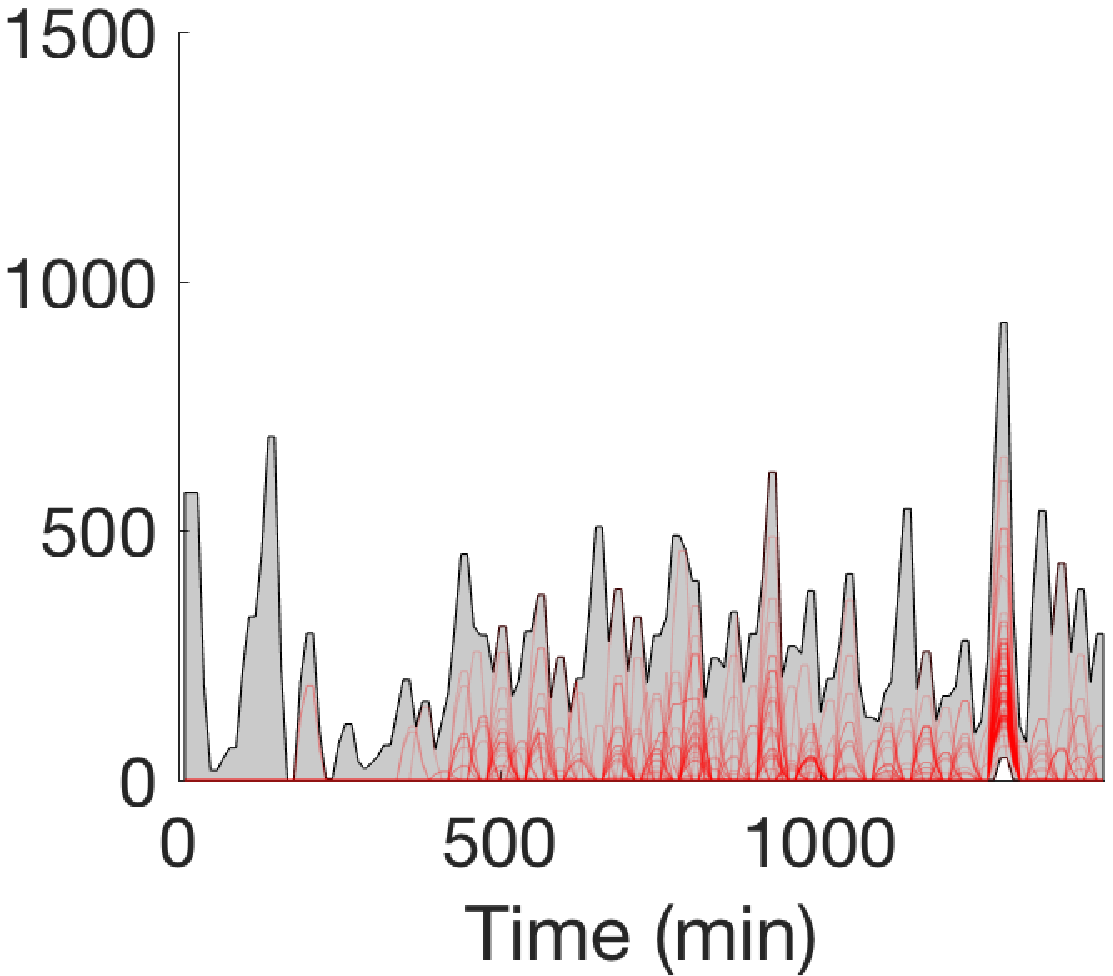}}
\hfill
\subfloat[Cluster \# 3: 592 people]{\includegraphics[height=\relheight\textwidth]{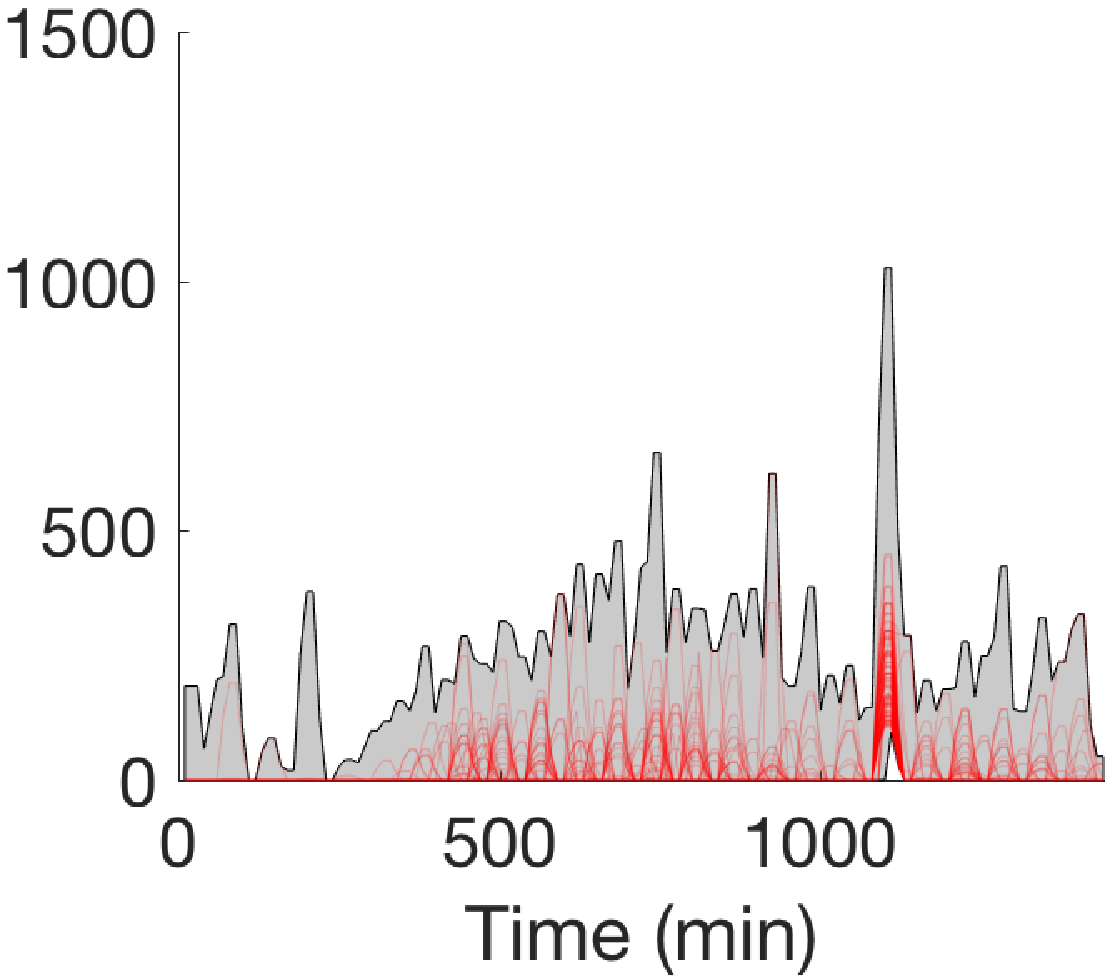}}
\hfill
\subfloat[Cluster \# 4: 440 people]{\includegraphics[height=\relheight\textwidth]{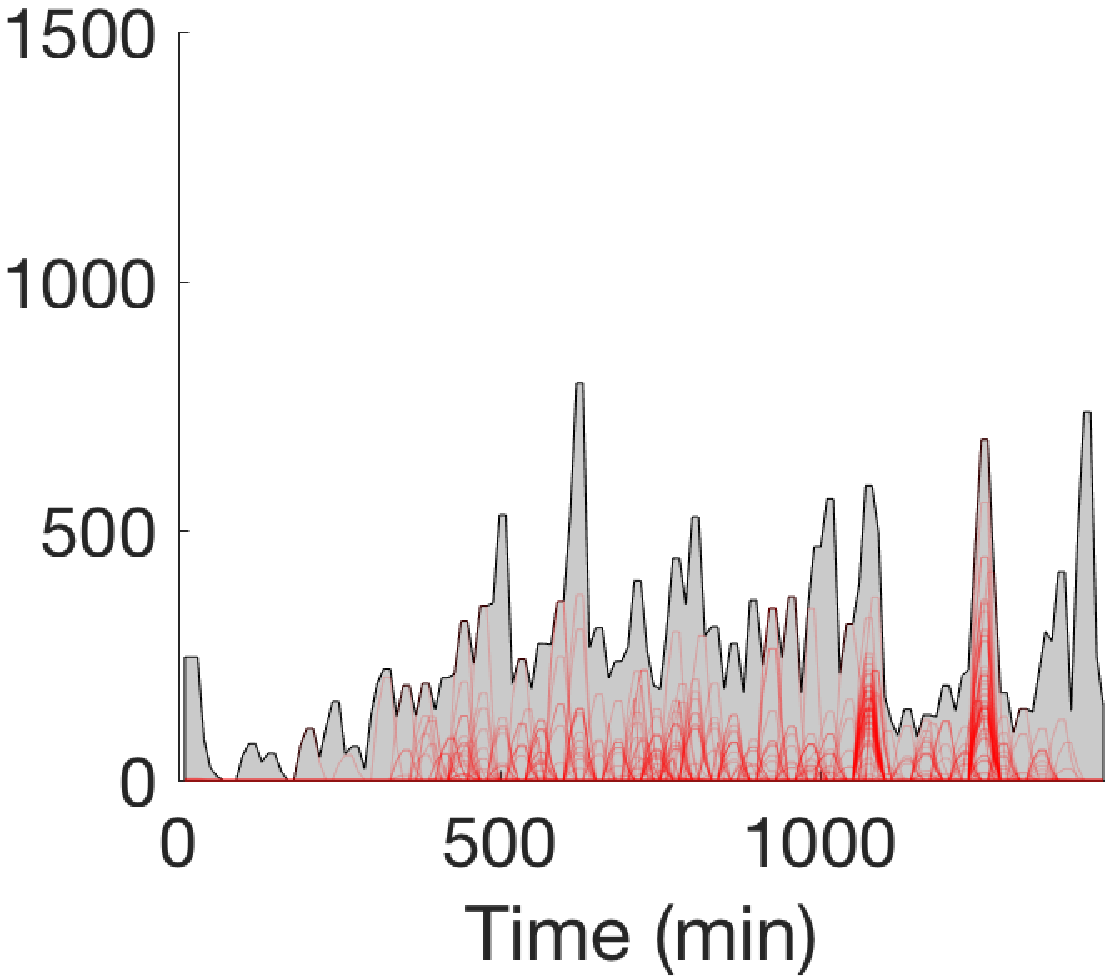}} \\ 
\subfloat[Cluster \# 5: 4663 people]{\includegraphics[height=\relheight\textwidth]{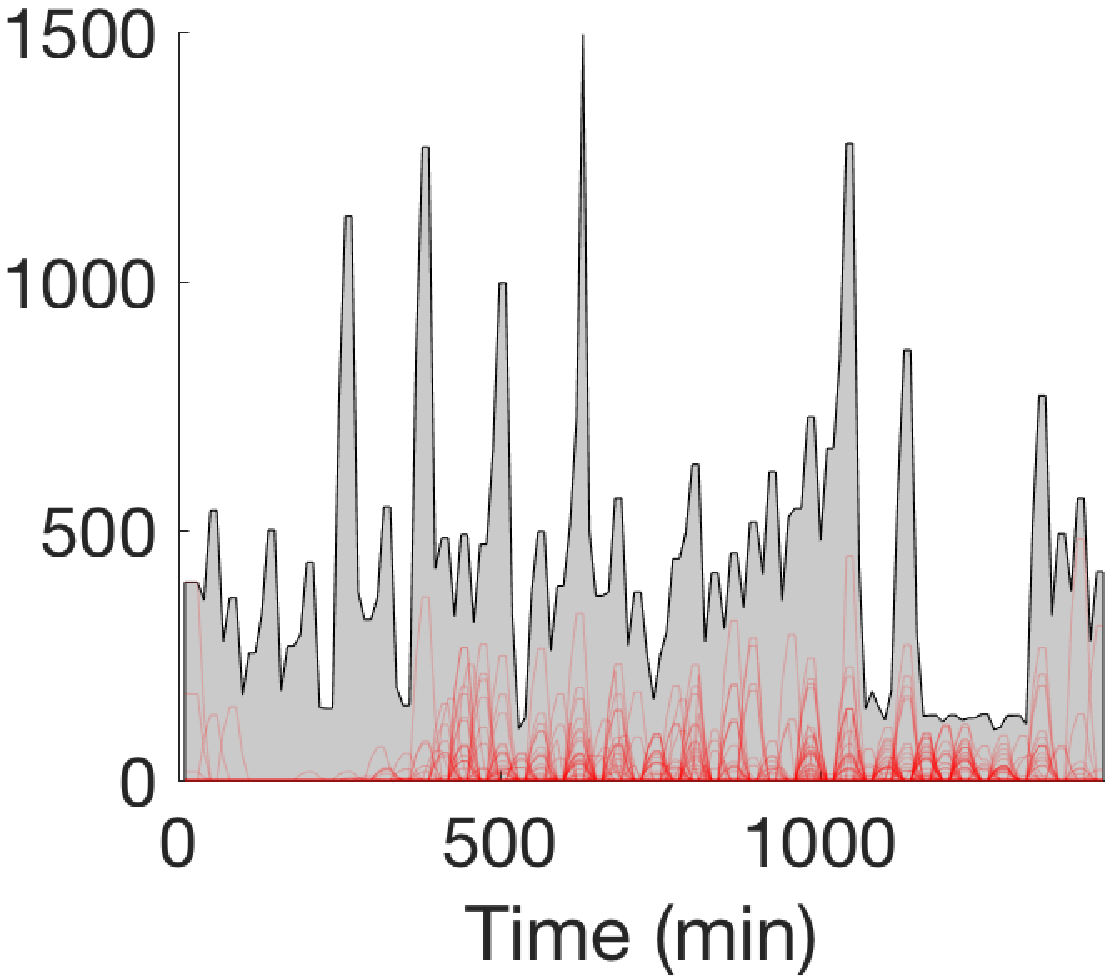}}
\hfill
\subfloat[Cluster \# 6: 325 people]{\includegraphics[height=\relheight\textwidth]{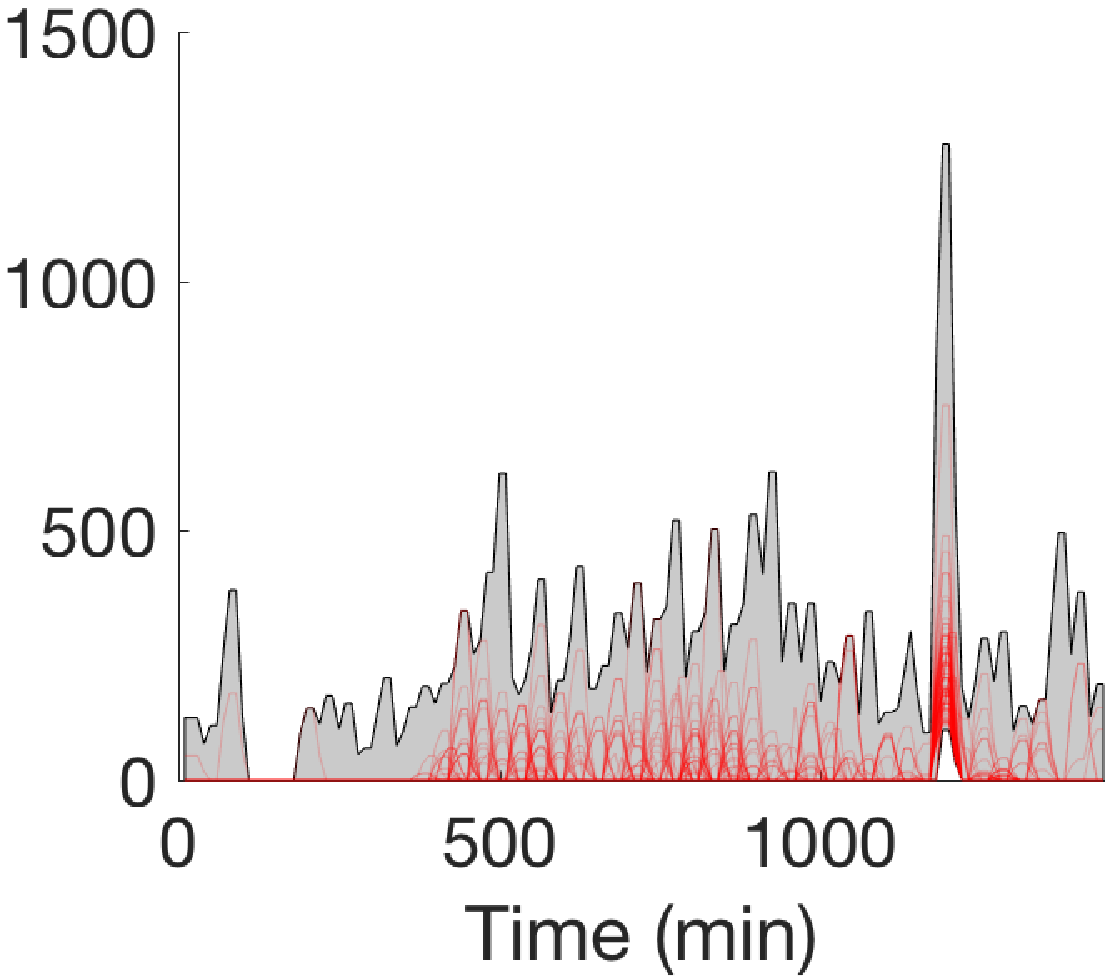}}
\hfill
\subfloat[Cluster \# 7: 662 people]{\includegraphics[height=\relheight\textwidth]{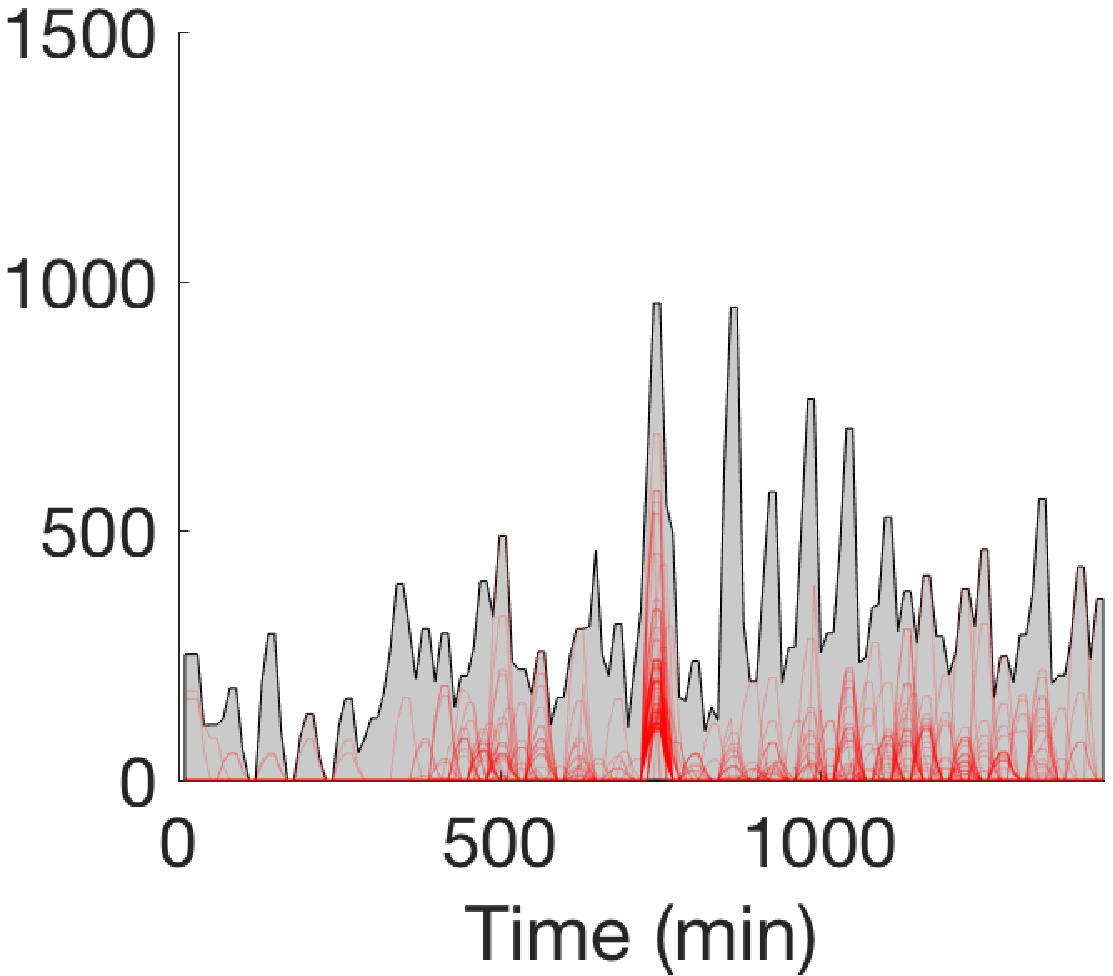}} \\ 
\subfloat[Cluster \# 8: 128 people]{\includegraphics[height=\relheight\textwidth]{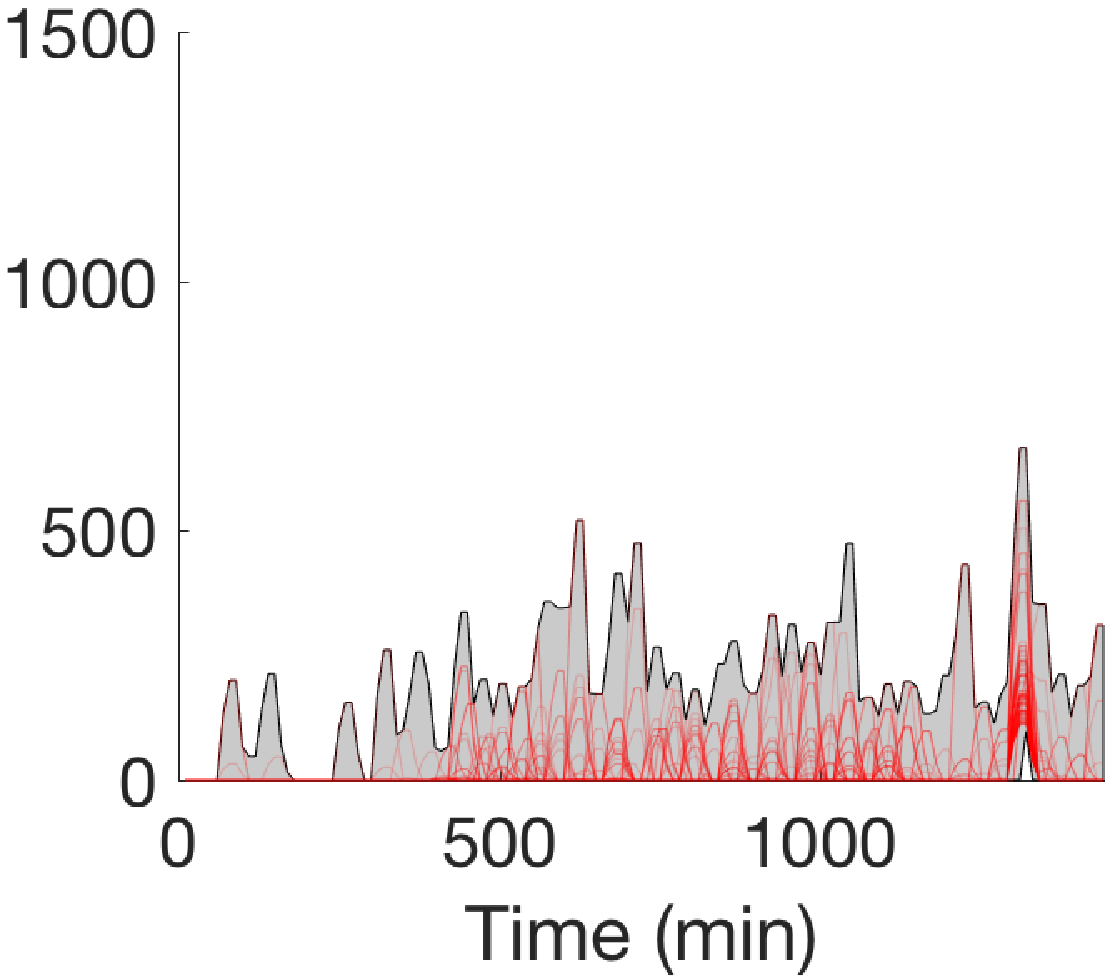}}
\hfill
\subfloat[Cluster \# 9: 658 people]{\includegraphics[height=\relheight\textwidth]{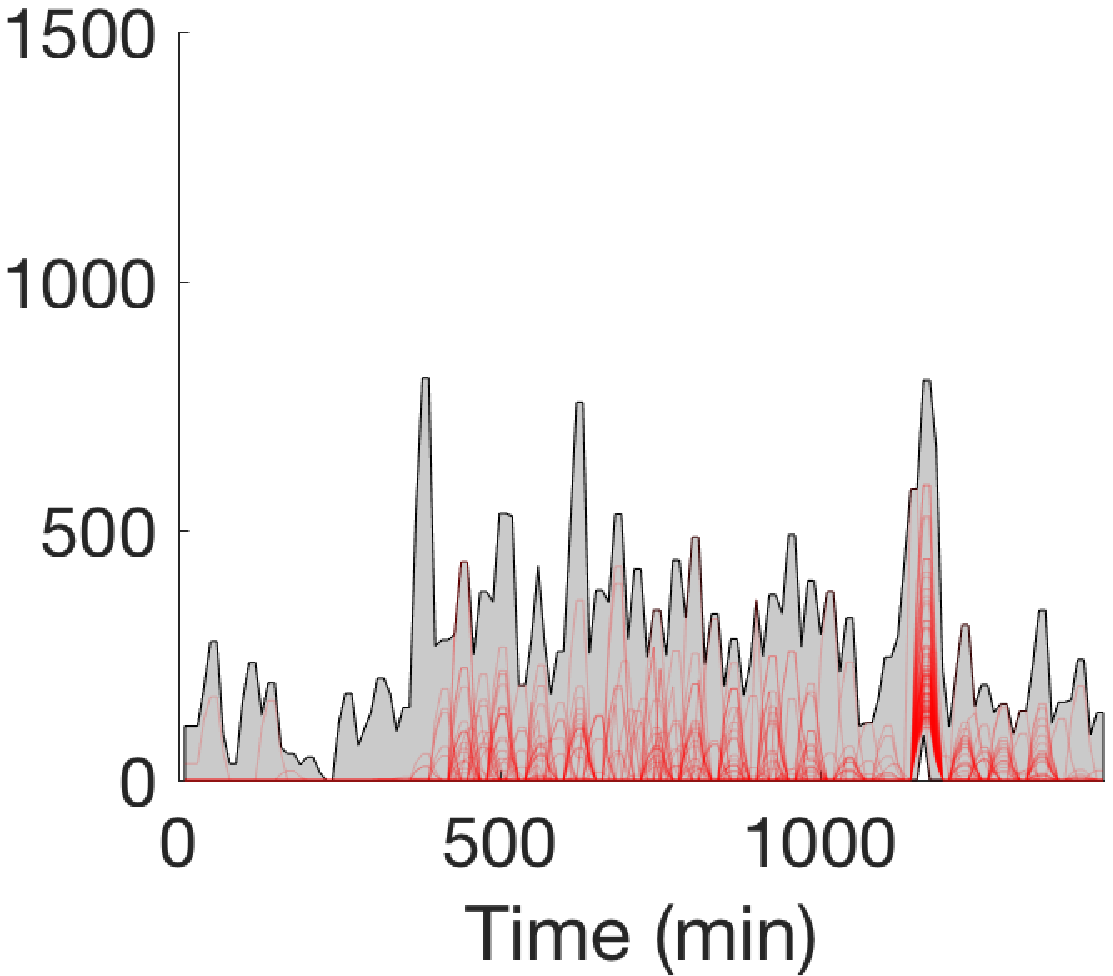}}
\hfill
\subfloat[Cluster \# 10: 551 people]{\includegraphics[height=\relheight\textwidth]{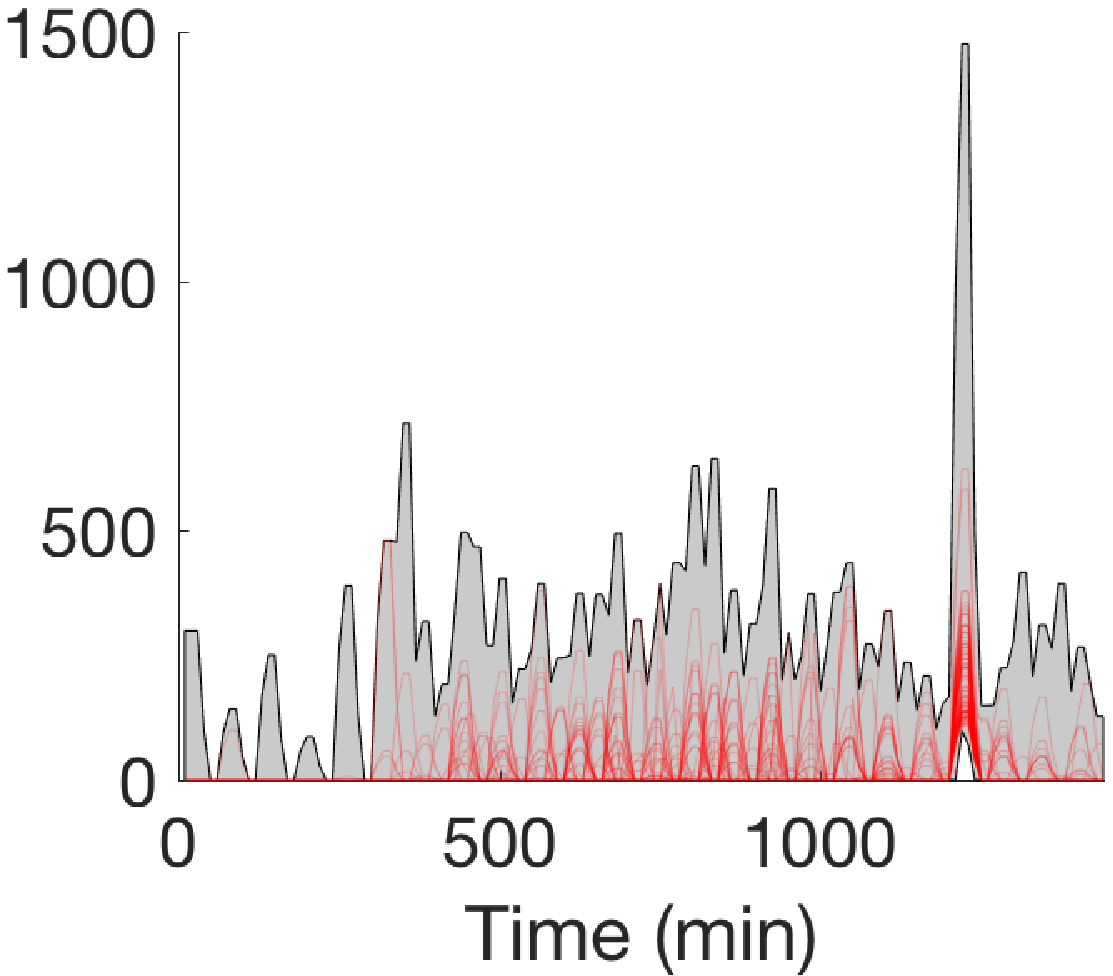}}
\caption{Uncertainty sets for clusters 2-10 extracted from the NHANES database. Cluster \# 1 was used for our experiments and is reported in Figure \ref{fig:nhanes_box}.  Cluster \# 2, 3, 4, 6, 8, 9, 10 show peaks at about minute 1200, indicating schedules characterized by CHO-rich dinners. Participant in cluster \# 7 are characterized by a rich lunch, while no particular patterns can be observed for cluster \# 5.}
\end{figure}

\end{document}